\documentclass[useAMS,usenatbib]{mn2e}

\usepackage{graphicx}
\usepackage{subfigure}
\usepackage{amssymb}
\usepackage{amsmath}
\usepackage{tabularx}
\usepackage{float}
\usepackage{placeins}

\title[Radiative Feedback and Cluster Formation]{Simulating radiative feedback and star cluster formation in GMCs: I. Dependence on gravitational boundedness}

\author[C.S.\ Howard, R.E.\ Pudritz, \& W.E.\ Harris ]{Corey \ S. \ Howard$^{1}$\thanks{E-mail: howardcs@mcmaster.ca}, Ralph \ E.\ Pudritz$^{1,2,3,4}$, William \ E.\ Harris$^{1}$\\
$^{1}$Department of Physics and Astronomy, McMaster University, 1280 Main St.~W, Hamilton, ON L8S 4M1, Canada\\
$^{2}$Origins Institute, McMaster University, 1280 Main St.~W, Hamilton, ON L8S 4M1, Canada \\
$^3$Zentrum f\"ur Astronomie der Universit\"at Heidelberg, Institut f\"ur Theoretische Astrophysik, Albert-Ueberle-Str. 2, 69120 Heidelberg, Germany \\
$^4$Max-Planck Institute f\"ur Astronomie, K\"onigstuhl 17, 69117 Heidelberg, Germany 
}

\begin{document}
\bibliographystyle{mn2e}

\date{16 June 2016}

\pagerange{\pageref{firstpage}--\pageref{lastpage}} \pubyear{2016}

\maketitle

\label{firstpage}

\begin{abstract}

Radiative feedback is an important consequence of cluster formation in Giant Molecular Clouds (GMCs) in which newly formed clusters heat and ionize their surrounding gas. The process of cluster formation, and the role of radiative feedback, has not been fully explored in different GMC environments. We present a suite of simulations which explore how the initial gravitational boundedness, and radiative feedback, affect cluster formation. We model the early evolution ($<$ 5 Myr) of turbulent, 10$^6$ M$_{\odot}$ clouds with virial parameters ranging from 0.5 to 5. To model cluster formation, we use cluster sink particles, coupled to a raytracing scheme, and a custom subgrid model which populates a cluster via sampling an IMF with an efficiency of 20\% per freefall time. We find that radiative feedback only decreases the cluster particle formation efficiency by a few percent. The initial virial parameter plays a much stronger role in limiting cluster formation, with a spread of cluster formation efficiencies of 37\% to 71\% for the most unbound to the most bound model. The total number of clusters increases while the maximum mass cluster decreases with an increasing initial virial parameter, resulting in steeper mass distributions. The star formation rates in our cluster particles are initially consistent with observations but rise to higher values at late times. This suggests that radiative feedback alone is not responsible for dispersing a GMC over the first 5 Myr of cluster formation.

\end{abstract}


\section{Introduction}

The modern paradigm for star cluster formation suggests that clusters are born from dense clumps (n $>$ 10$^4$ cm$^{-3}$) which form in supersonically turbulent, filamentary molecular clouds \citep{Lada2003,MaclowKlessen,BertoldiMckee,Kruijssen2012}. Recent observations show that young stellar clusters tend to form at the intersection of filaments in regions that are fed by higher than average accretion rates \citep{Schneider2012,Kirk2013,Balsara,Banerjee2006}. These overdense regions may then fragment further, resulting in highly subclustered objects which undergo mergers \citep{Megeath2012,Kuhn2012}. 

The question of how the high accretion flow onto dense, star-forming clumps is halted is of particular importance. The mass accretion history of these objects has implications for the final cluster mass and, hence, the observed cluster mass function. Moreover, the overall conversion of molecular gas to stars is inefficient \citep{Lada2003,Murray2011}. Understanding what processes limit the star formation efficiency in molecular clouds is key to a complete star formation theory as well as underpinning all discussions of simulations and theories of galaxy formation and evolution.

Several mechanisms have been proposed to explain the low star formation efficiency observed in molecular clouds. For example, turbulent velocity fields have been shown to lower the overall star formation rate per freefall time \citep{Klessen2000,Bate2003,Bonnell2008}. This cannot be the sole mechanism at work, however, since, given enough time, all of the gas will be converted to stars. The added pressure support via magnetic fields can also lower the overall star formation efficiency \citep{MyersGoodman1988,Tilley2007,Federrath2012}. While magnetic fields may play an important role in limiting cloud fragmentation, they also contribute to the dispersal of gas within individual star forming cores by means of hydromagnetic bipolar outflows and jets that are associated with young stars of all masses \citep{Matzner,Federrath2014}.

The process of feedback from newly-formed stars can both limit the star formation efficiency and disperse the surrounding gas, thereby halting the star formation process. Feedback comes in various forms: protostellar jets \citep{Li2006,Maury2009,Federrath2014}, stellar winds \citep{Dale2008}, ionization and heating of the gas \citep{Dale2005,Peters2010,Klassen2012}, and radiation pressure \citep{Krumholz2012,Murray2010}. Radiative feedback has been suggested as being the most important form of feedback in clusters which host massive star formation \citep{Murray2010,Dale2012,Bate2012}. The ionizing radiation released from newly-formed stars heats the gas to approximately 10,000 K and can drive the expansion of massive HII regions. The direct input of momentum via high energy photons interacting with surrounding dust grains can also drive strong outflows.

Previous studies of radiative feedback on both small and large spatial scales have shown that the star formation efficiency can be significantly reduced \citep{Dale2007,Peters2010,Dale2012,Bate2012,Klassen2012,Kim2012}. Studies which model the formation of individual stars, however, are typically limited to the clump scale or low mass molecular clouds. These studies therefore neglect the impact of radiative feedback on the global structure of massive GMC's. Since cluster formation can be broadly distributed throughout a cloud, cluster interactions over the entire range of the GMC must be considered. There have been studies of radiative feedback in massive GMCs (10$^6$ M$_{\odot}$) \citep{Dale2012}. Due to numerical constraints, the supergiant molecular cloud regime ($>$ 10$^6$ M$_{\odot}$) has not been examined nor do most models follow the cloud's evolution up to the point of dispersal which is crucial to placing constraints on the lifetime of molecular clouds. Recent results suggest that radiation feedback 
alone is not effective in accounting for low star formation rates in galaxy formation simulations \citep{Agetz}.

The lifetimes of molecular clouds, and the processes responsible for destroying them, are challenging to measure. Estimates of GMC lifetimes range from a single freefall time \citep{Elmegreen2000,Hartmann2001}, $\tau_{ff}$, up to tens of freefall times \citep{Scoville1979,Scoville2004}. Recent observations of the deuterium fraction in both massive, starless cores and Infrared Dark Clouds (IRDCs) indicates ages within a range of 6 to 10 local freefall times \citep{Kong2015,Barnes2016}, corresponding to $\sim$3 Myr for IRDCs. We note that direct measurements of the lifetimes of molecular clouds out to a redshift of 4 are now possible with ALMA (Atacama Large Millimeter Array) and the method outlined in \citet{Kruijssen2014}. Theoretical estimates suggest that radiative feedback, in particular radiation pressure, is the most important physical process responsible for destroying massive GMC's \citep{Murray2010} but numerical simulations of star formation in GMC's which include radiative feedback are required to validate these claims.   

The role of radiative feedback in different molecular cloud environments has also not been explored in detail. The observed clouds in the Milky Way and extragalactic sources show a wide range of properties. More specifically, molecular clouds have been observed to have a range of virial parameters, defined as the ratio of the cloud's kinetic energy to gravitational potential energy. This ratio may vary from 0.5 up to 10 with the mean being approximately 1 \citep{Solomon1987,Rosolowsky2007,Hernandez2015}. Simulations have shown that dense, self-gravitating clumps can still fragment out of clouds that are globally unbound \citep{Clark2005,Clark2008,Bonnell2011}. Furthermore, the star formation efficiency is naturally lowered in unbound clouds \citep{PPadoan2012}. Radiative feedback may play a stronger role in lowering the star formation efficiency in unbound versus bound clouds since the added energy and momentum can more easily disperse the surrounding, unbound gas. 

To investigate these questions, we have chosen to simulate the formation of young stellar clusters in molecular clouds with a range of properties while including detailed radiative feedback. As described above, we focus on varying the initial virial parameter in turbulent, 10$^6$ M$_{\odot}$ GMCs. We are focusing on this mass because the majority of star formation in the Milky Way is hosted in these clouds \citep{Murray2011}. 

In order to simulate the long term evolution of these clouds, we use sink particles to represent young clusters and combine this with a subgrid model to track star formation within the clusters. In Section 2, we describe the details of this subgrid model and the initial conditions used in our simulations. In Section 3, we describe the evolution of clouds with varying initial virial parameters, discuss the role of radiative feedback, and compare cluster properties across clouds. We find that radiative feedback has only a slight effect on the star formation efficiency.  We find, in fact, that efficiencies are dominated by the gravitational binding of the cloud and that virial and sub virial initial cloud models do not show the low efficiencies that observations demand. In Section 4, we discuss how our results compare to recent observations of local star forming regions, followed by our conclusions in the final section.

\section{Numerical Methods} \label{methods}

We perform numerical simulations using version 2.5 of the hydrodynamical code FLASH \citep{Fryxell2000}. FLASH is an adaptive mesh refinement code (AMR) which 
integrates the compressible gas dynamic equations on a Eulerian grid and includes modules to treat self-gravity, radiative transfer, and star formation. We include cooling via molecular lines and dust \citep{Banerjee+2006}, and use a simple ideal gas law equation of state with an adiabatic index of 1.67.

Radiative transfer is handled via a hybrid-characteristics raytracing scheme developed by \citet{Rijkhorst} and adapted for star formation simulations by
\citet{Peters2010}. This scheme follows the propagation of both ionizing and non-ionizing radiation and makes use of the DORIC routines \citep{Frank1994,Mellema2002}. The DORIC package is an iterative scheme used to calculate the ionization, heating, and cooling rates of a large number of species \citep{Frank1994}. For simplicity, however, hydrogen is used as the only gas component. \citet{Peters2010} have shown that the code can accurately reproduce the analytic solution for the expansion of a D-type ionization front from \citet{Spizter1978}, and \citet{Iliev2006} have shown the code accurately treats R-type fronts. The opacity to the non-ionizing radiation is represented by the Planck mean opacities from the dust model of \citet{Pollack1994}. The Planck mean opacities are used because the raytracer has no frequency dependence apart from ionizing versus non-ionizing radiation. Radiation pressure is included by adopting a single UV opacity of 775 cm$^2$ g$^{-1}$ from \citet{LiDraine2001}. The associated 
radiative force per unit mass is given by,

\begin{equation}
  F = \frac{L}{c}\frac{e^{-\tau}}{4\pi r^2}
\end{equation}

\noindent where $L$ is the source luminosity, $c$ is the speed of light, and $\tau$ and $r$ are the optical depth and distance between the source and the target cell.

To represent the formation of star clusters, we make use of Lagragian sink particles (Banerjee et al. 2009; Federrath et al. 2010). In order to form a particle, a region
of gas in the simulation volume must meet the following conditions:

\begin{itemize}
  \item At the highest level of refinement
  \item The divergence is less than zero (ie. converging)
  \item At a local gravitational minimum
  \item The region is Jeans unstable
  \item The region is gravitationally bound
  \item Not within the radius of another particle.
\end{itemize}

These particles were designed to represent stars but the above conditions also model the formation of a dense clump which hosts cluster formation.

In order to examine the impact of radiative feedback from stellar clusters, we require a subgrid model for the radiative output of a cluster as it evolves over time. Below, we provide a brief summary of this model. A more detailed description can be found in Howard et al. (2014).

\subsection{Subgrid model for cluster formation: Cluster sinks}

\indent One of the most important aspects of radiative feedback of a young forming cluster on its surrounding host GMC is the shutting off of the accretion flow into the cluster forming region. A cluster must therefore be assigned the correct, combined radiative output of all its member stars as star formation proceeds. Here, we provide a brief description of the subgrid model we use to form stars within a cluster particle. 

In order for cluster formation to begin, the host clump must reach sufficient density in order to fragment and collapse. The threshold density for cluster formation has been observationally measured by \citet{Lada2003} who found that the transition from starless to star forming clumps, in local star forming regions, occurs at a density of $\sim$10$^{4}$ cm$^{-3}$. This number has also been quoted by other authors \citep{Lombardi2010,Heidermann2010} who found that the SFRs in molecular clouds are well correlated to the dense ($>$10$^4$ cm$^{-3}$) gas mass. More recent work by \citet{Kainulainen2014}, aimed at producing probability density functions for gas density from column density observations, found a density threshold of 5$\times$10$^3$ cm$^{-3}$. Analytical models for the density threshold, which also account for environmental effects, find a threshold of 1.5$\times$10$^{4}$ cm$^{-3}$ when adopting Solar neighbourhood parameters \citep{KrumKee2005,PadNord2011,Kruij14}. We have therefore chosen to use a fiducial density threshold for cluster formation of 10$^4$ cm$^{-3}$. We report on several simulations in which we vary this value by an order of magnitude.

Once this critical density threshold is reached, as well as the other conditions outlined above, a cluster, represented by a sink particle in this work, forms. Star formation then proceeds within this cluster as gas accretes onto this dense region.

\indent With this general framework in mind, we use our subgrid model for the formation, radiative feedback, and evolution of a cluster that addresses how the original gas reservoir (ie. clump) is divided into stars over time and how accreted gas is handled \citep{Howard2014}. We refer the reader to that paper for the details, which we now briefly summarize.  

\indent To treat the conversion of reservoir mass into stars, we divide it into main sequence stars at a prescribed efficiency according to an IMF. We use a star formation
efficiency of 20$\%$ per freefall time \citep{Lada2003}, where the freefall time is given by

\begin{equation}
t_{ff} = \sqrt{\frac{3\pi}{32 G n \mu m_{H}}}
\end{equation}

\noindent where $n$ is the number density, $\mu$ is the mean molecular weight, and $m_{H}$ is the mass of hydrogen. As discussed above, our fiducial value for $n$ is taken to be 10$^{4}$ cm$^{-3}$. The resulting freefall time is 0.36 Myr.

\indent We use the Chabrier IMF, given by

\begin{equation}
\xi(log \ m) = 
\begin{cases} 
0.093\times exp\{\frac{-(log \ m - log \ 0.2)^2}{2\times(0.55)^2}\}, & \mbox{\ m $\leq$ 1 \ M$_{\odot}$ } \\
0.041m^{-1.35 \pm 0.3}, & \mbox{\ m $>$ 1 \ M$_{\odot}$},
\end{cases}
\end{equation}
\newline 

 \noindent as the probability distribution from which we draw our stars \citep{Chabrier2005}. We chose to sample the Chabrier IMF every tenth of a freefall time, or 3.6$\times$10$^4$ years. As shown in \cite{Howard2014}, this sampling frequency allows cluster properties to evolve smoothly over time while still accurately reproducing the IMF from which the stars are drawn.     

The masses of all stars formed in a cluster are recorded, meaning that its stellar content is known at all times. Using the analytic fits for the luminosity of main sequence stars provided by \citet{Tout1996}, we know the luminosity of each star within the cluster. The total luminosity, and ionizing photon rate, of each cluster is the sum of the stellar components. These quantities are then used by the raytracer to treat the radiative transfer.

Gas accreted by a cluster is simply added to the gaseous reservoir. Since we are sampling an IMF in order to get our final stellar masses, we do not need to treat accretion onto individual stars. The accreted gas can then be used to form new stars at a later time. 

We do not include mass loss from the cluster particle because we assume that the reservoir gas remains gravitationally bound to the cluster over long timescales, even in the presence of stellar feedback. Earlier simulations by \cite{Dale2005,Dale2007} found that the inclusion of radiative feedback was $not$ sufficient to unbind the majority of the clump mass. The reason for this is collimated ionized outflows which are released into low density regions perpendicular to the dense filaments out of which the stars form. We therefore assume our gas reservoir is in a similarly filamentary state, preventing its disruption via stellar feedback. Furthermore, recent numerical works \citep{KrujMasch12,DaleErcolano2015}, both with and without feedback mechanisms, and observational studies \citep{Ginsburg2016} suggest that cluster formation proceeds via gas exhaustion rather than gas expulsion. This further justifies the use of cluster particles which do not lose mass. 

We note that, when performing a simulation, we apply a mass threshold below which particles do not radiate. This was implemented in order to reduce the computational time. The raytracing scheme is computationally expensive so by allowing
only massive particles to radiate, the total time spent during the raytracing step can be greatly reduced. We have chosen a threshold of 1000 M$_{\odot}$ contained in \emph{stars}. Note
that this is not the \emph{total} mass of the cluster particle, but rather the stellar mass. We have chosen this threshold because it is approximately the mass at which the first O stars are expected to appear \citep{Howard2014}. We have verified that, at any given time, particles above this mass dominate
the total luminosity budget so we are therefore confident that the application of a mass threshold should not impact our results significantly.

We have performed tests of the raytracing scheme similar to \citet{Peters2010}, except with our cluster sink particles, as described at the beginning of this Section. These were low resolution simulations, with 128 grid cells along each simulation axis, where a cluster sink particle was placed at the center of a uniform medium with a number density of 10$^3$ cm$^{-3}$. We allowed the cluster particle to sample the IMF once at the beginning of the simulation, after which no new stars were added to the cluster, and compared the resulting HII region to the analytical expression for the expansion of a D-type ionization front from \citet{Spizter1978}. We find that, during expansion, the size of the HII region does not differ by more than 15\% from the analytical solution. This error is reduced to less than 9\% if we instead compare to the analytical expression presented in \citet{Raga2012}, as was done in the radiative transfer code comparison project \emph{STARBENCH} \citep{Starbench}.

The addition of radiation pressure produces an HII region that is $\sim$10\% larger compared to the test without radiation pressure. Additionally, the central density inside the HII region is significantly lower when radiation pressure is 
included, differing by approximately two orders of magnitude compared to the test without radiation pressure. 

\subsubsection{Cluster Sink Mergers}

 \begin{figure}
 \begin{center}
  \includegraphics[width=0.9\linewidth]{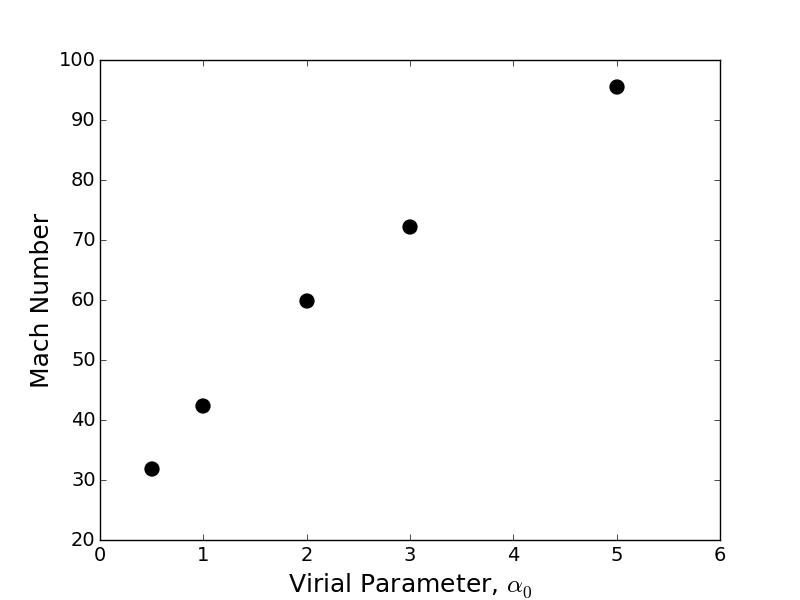}
 \end{center}
 \caption{The Mach numbers corresponding to the initial virial parameters.}
 \label{fig:mach}
\end{figure}

The sink particle routines in FLASH allow for particle mergers \citep{Federrath2010}. In order for a merger to occur, several conditions must be met. Firstly, the particles must be closer than an accretion radius, defined to be 2.5 cells at the highest level of refinement. Secondly, the particles must be approaching one another. Lastly, the particles must be gravitationally bound to one another. More specifically, the total gravitational potential energy of the particle pair, at a distance of an accretion radius, must be greater than the total kinetic energy. When all of these conditions are met, the smaller mass particle is merged to the more massive particle. The new position is taken to be the center of mass position and the resulting particle velocity is determined via momentum conservation. 

We have made custom modifications to the merging routines in order to better represent the merging of cluster sink particles. As mentioned above, cluster sink particles have their mass divided between fully-formed stars and the gaseous reservoir. When a cluster particle merger occurs, we combine the reservoir masses and the total stellar mass. Since we track the masses of all stars formed in each particle, the total stellar content remains the same after merging. We emphasize that, while we will be using merger to describe the joining of two cluster sink particles, this differs from the dynamical mergers associated with fully formed clusters that are devoid of gas. In our context, a merger is more aptly defined as the coalescence of a central stellar cluster as well as the envelope of surrounding gas.

Recent studies indicate that gas poor clusters may form before feedback starts acting since the local freefall time in these dense regions is the highest \citep{KrujMasch12,DaleErcolano2015}. This means that the stellar cluster within our particles may be spatially segregated from the surrounding gas reservoir. Therefore, the merger of the stellar component in our particles may be thought of in the traditional way, namely a dynamical merger of gas poor stellar clusters. The reservoir can be interpreted as the cocoon of gas which surrounds the central cluster, which is also joined during the merger. We do not resolve the structure of the gas reservoir, however, so we cannot distinguish between the merger of gas rich clusters and gas poor clusters with a surrounding envelope of gas.

\begin{figure*}
\begin{tabular}{c}
\includegraphics[width=0.8\linewidth]{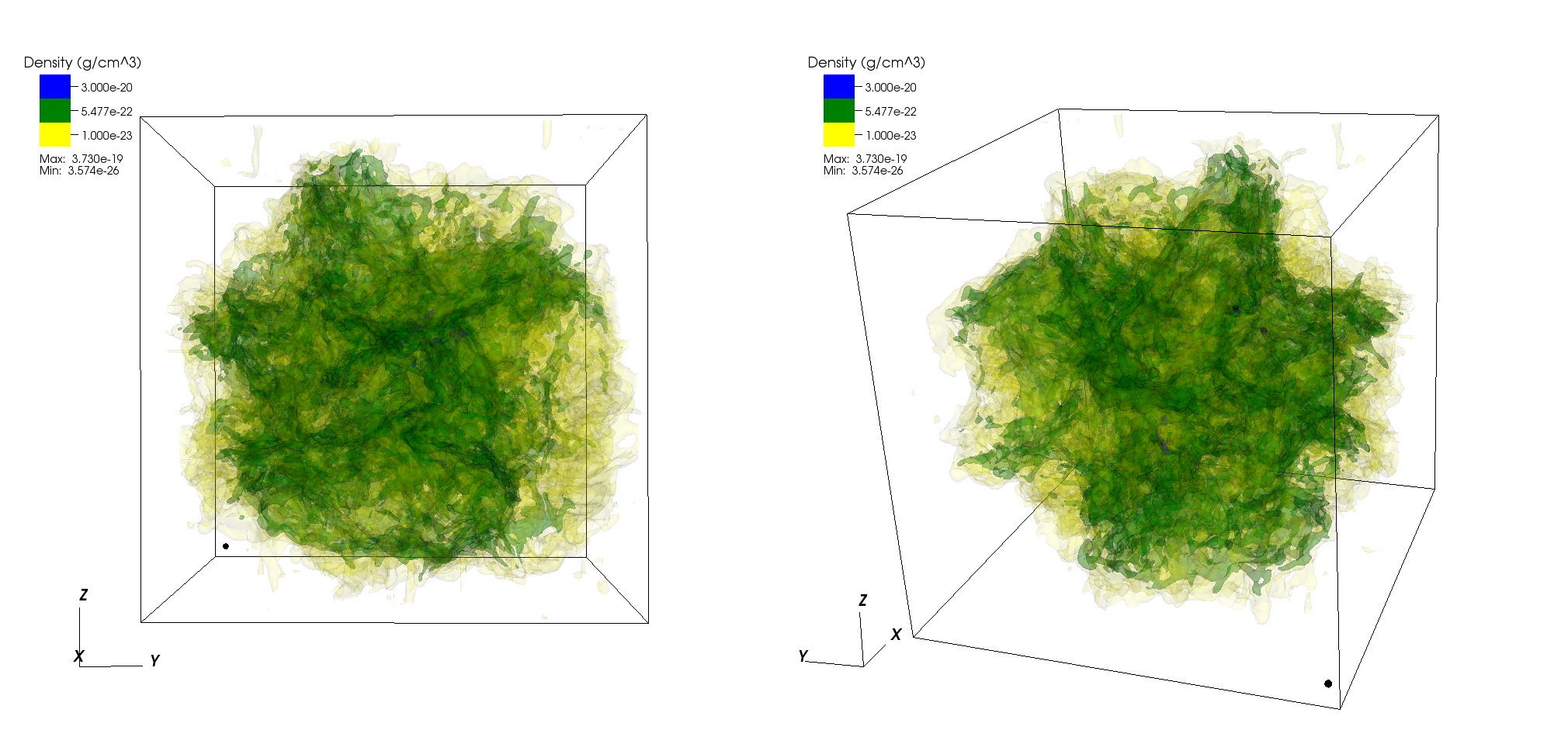} \\
\includegraphics[width=0.8\linewidth]{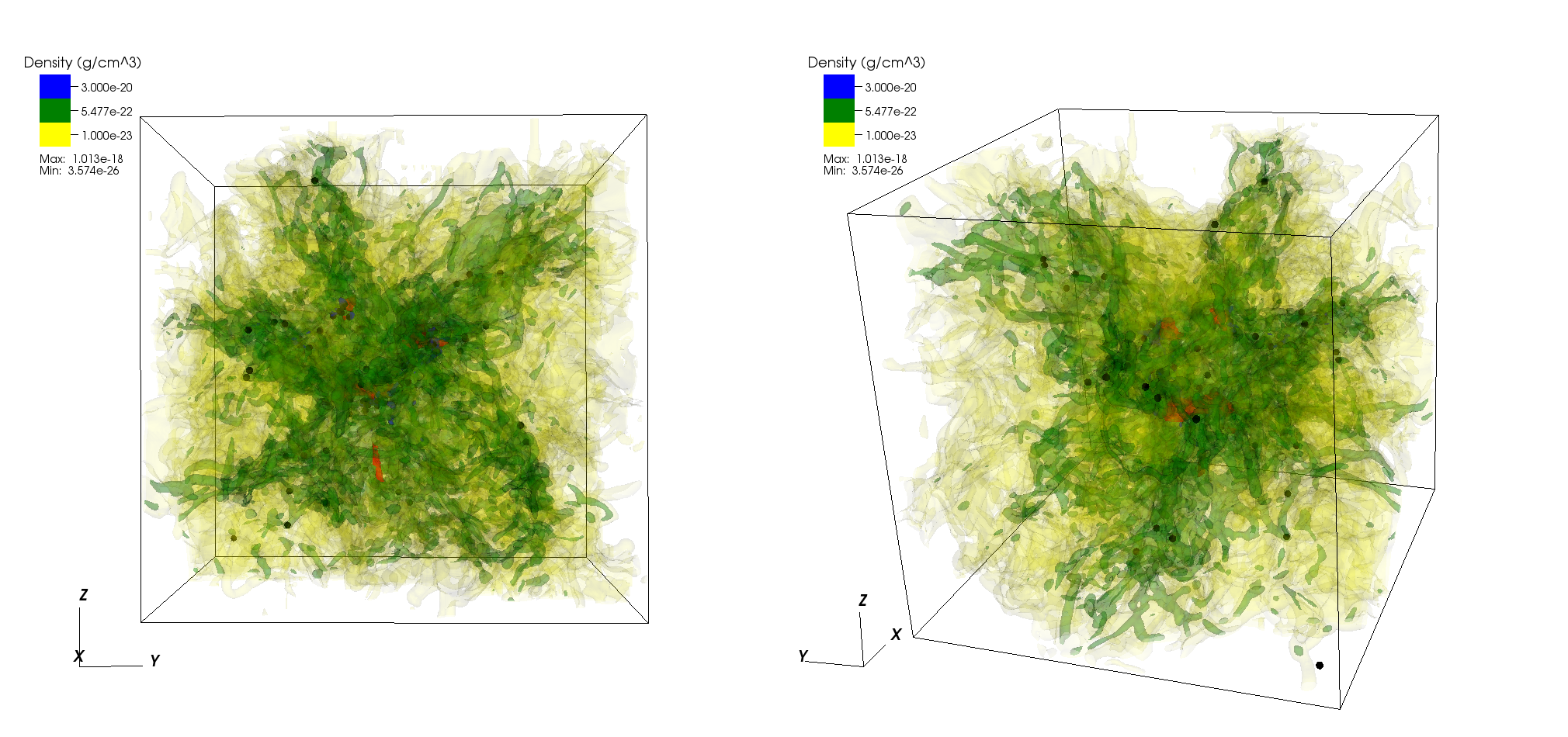} \\
\includegraphics[width=0.8\linewidth]{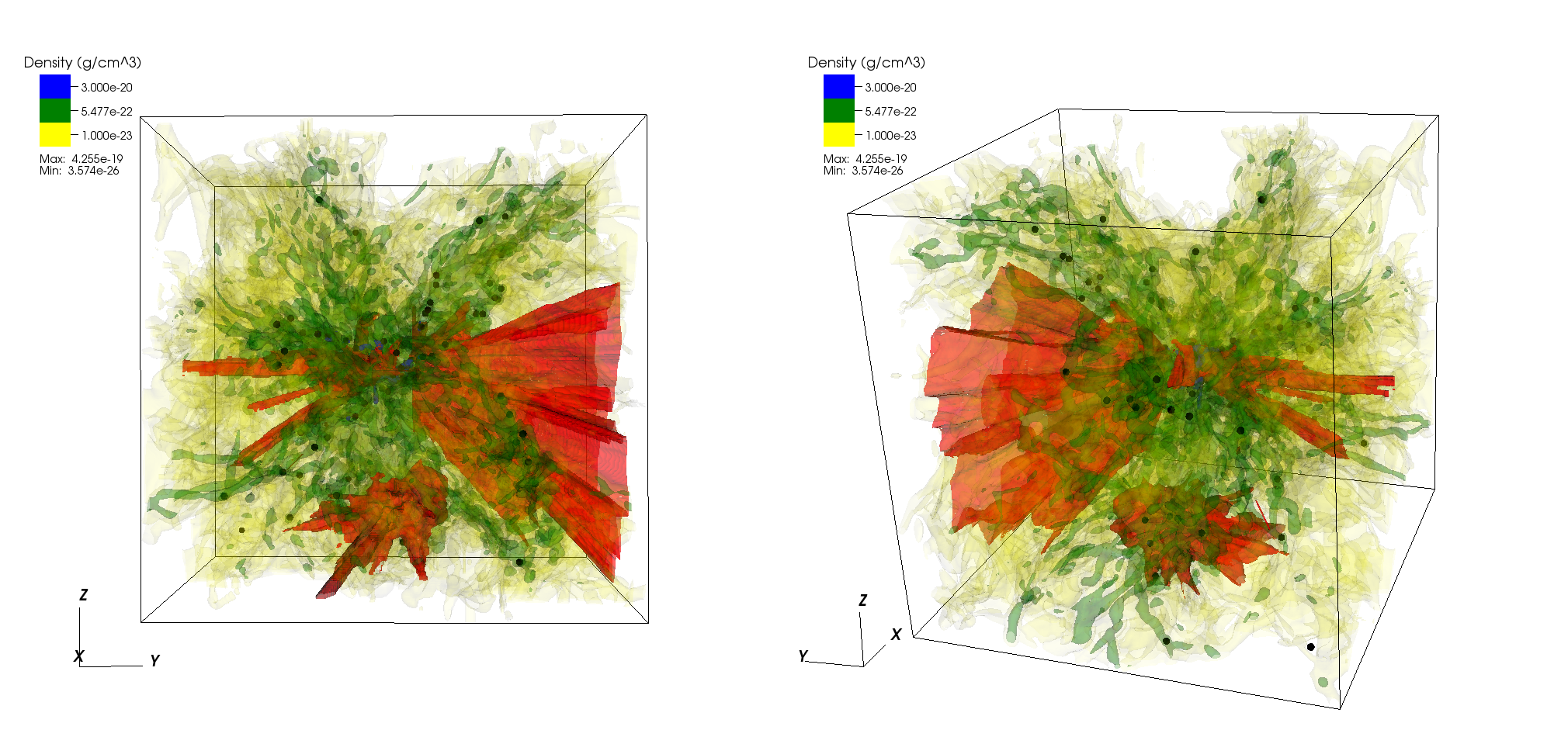} \\
\end{tabular}
\caption{A three-dimensional view of the $\alpha_0$ $=$ 1 simulation. The density contours are shown in yellow, green, and blue and the ionized regions are shown in red. The black 
circles represent cluster particles. Rows show two different views at the same time and columns show the state of the simulation at 0.97, 2.68, and 3.65 Myr from top to bottom. }
\label{fig:3d}
\end{figure*}

\subsection{Initial Conditions}

We simulate the formation of stellar subclusters in collapsing GMC's with different initial properties and examine the role of radiative feedback in these environments. The clouds have the same initial average of $n = 100$ cm$^{-3}$ and masses of 10$^6$ M$_{\odot}$. The radius of all clouds is 33.8 pc, corresponding to a minimum cell size of 0.13 pc. This results in a particle accretion radius of 0.325 pc (ie. 2.5 cells at the highest level of refinement). We have chosen an initial density profile which is uniform in the inner half of the cloud and decreasing as r$^{-3/2}$ in the outer half of the cloud. A quadratic fit is applied in the transition from uniform to decreasing density to ensure a continuous and smooth function.  

We use outflow conditions at the boundary of the simulation volume. The total mass in our simulations is therefore not conserved but can decrease over time as gas escapes the domain. This is particularly relevant to the discussion in the next Section.

The cloud is overlaid with a Kolomogorov turbulent velocity spectrum. The turbulence is not driven after it is initially imposed and its strength is determined via the virial parameter.

The virial parameter relates the total kinetic energy to the total gravitational potential energy in a cloud. It is defined as \citep{BertoldiMckee},

\begin{equation} \label{virial}
\alpha = 2\frac{E_{kin}}{|E_{grav}|} \approx \frac{5 \sigma^2 R}{G M}
\end{equation}

\noindent where $E_{kin}$ is the total kinetic energy, $E_{grav}$ is the total gravitational potential energy, $\sigma$ is the velocity dispersion, $R$ is the cloud radius, and $M$ is the cloud mass. Equation \ref{virial} is exact for a uniform density cloud. The velocity dispersion can be expressed as $\sigma = M_{a}\times c_s$ where $M_a$ is the Mach Number and $c_s$ is the sound speed, which is taken to be 2.3$\times$10$^5$ cm/s at 10 K. For a uniform density cloud without magnetic fields, a virial parameter of 1 corresponds to virial equilibrium and $\alpha > 2$ represents gravitationally unbound clouds. We use the first relation in Equation \ref{virial} since the kinetic and gravitational potential energies are easily measurable in our simulations.

We have chosen to examine initial virial parameters, defined as $\alpha_0$ henceforth, ranging from 0.5 to 5, which covers both bound, virialized, and unbound clouds. This range of virial parameters is within the range of observed GMC's \citep{Solomon1987,Rosolowsky2007,Hernandez2015}.  The corresponding Mach Numbers can be found in Figure \ref{fig:mach}. 

The models are evolved to $\sim$5 Myr, at which point supernovae are expected to inject large amounts of energy into the surrounding gas \citep{Starburst99}. Since the physics of Supernovae are not handled in the present code, we do not evolve the models past this point. 

\section{Results}

\subsection{Global Evolution and Gas Properties}

\begin{figure*}
\begin{tabular}{c}
\includegraphics[width=0.8\linewidth]{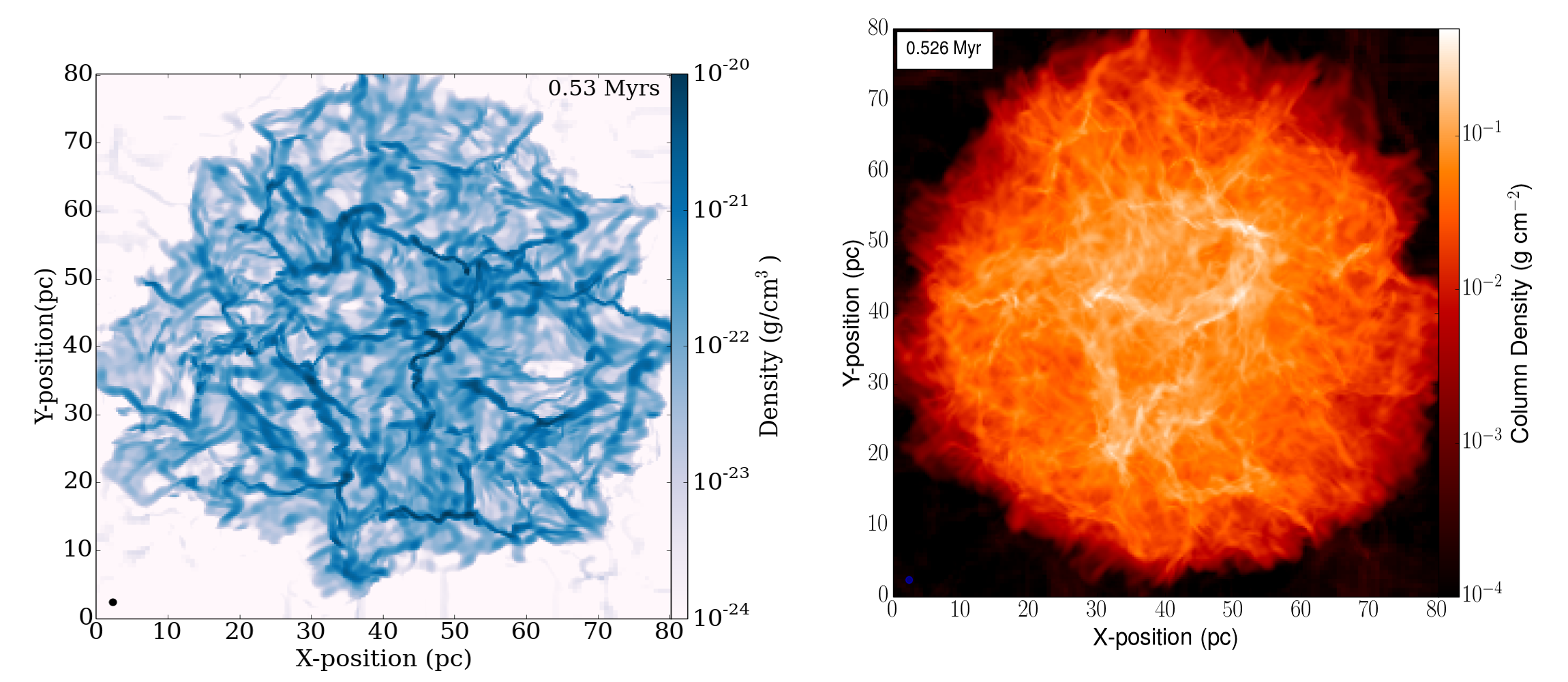} \\
\includegraphics[width=0.8\linewidth]{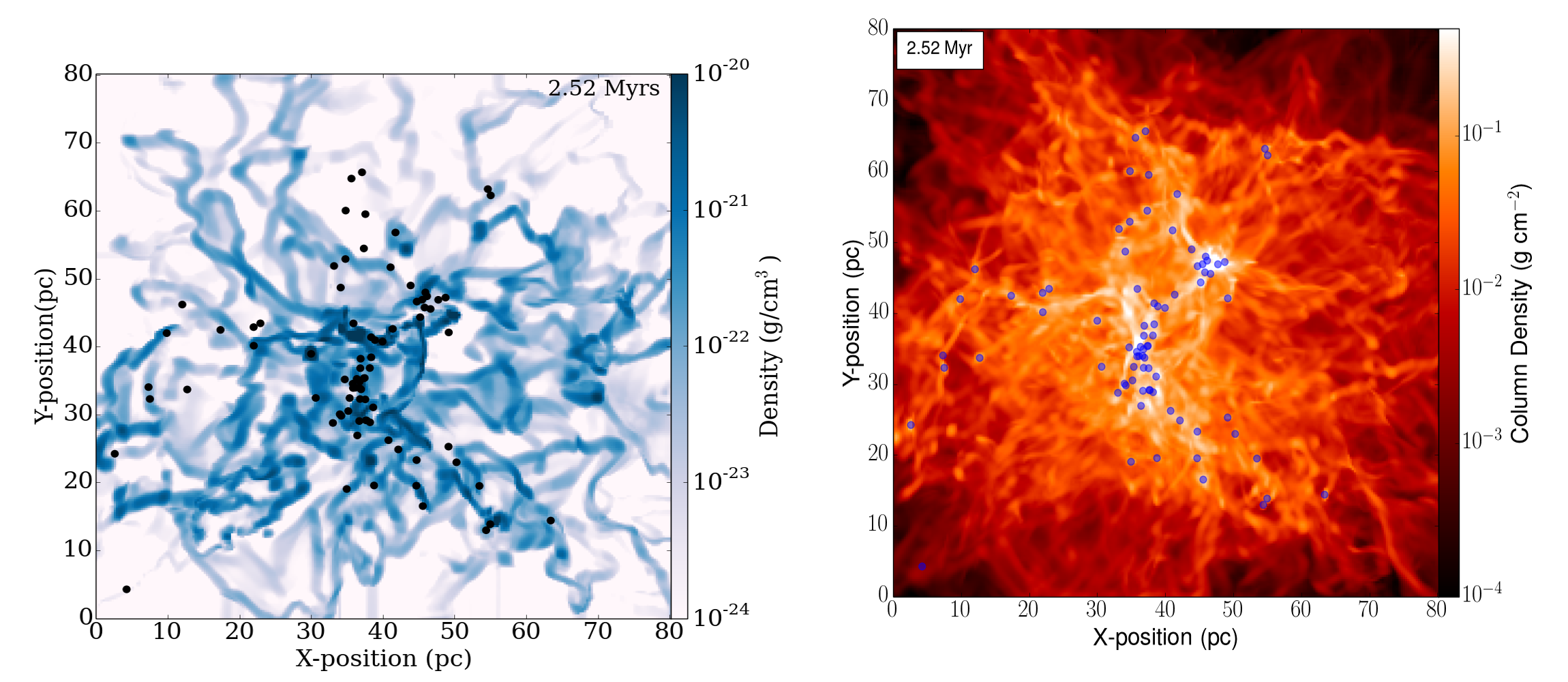} \\
\includegraphics[width=0.8\linewidth]{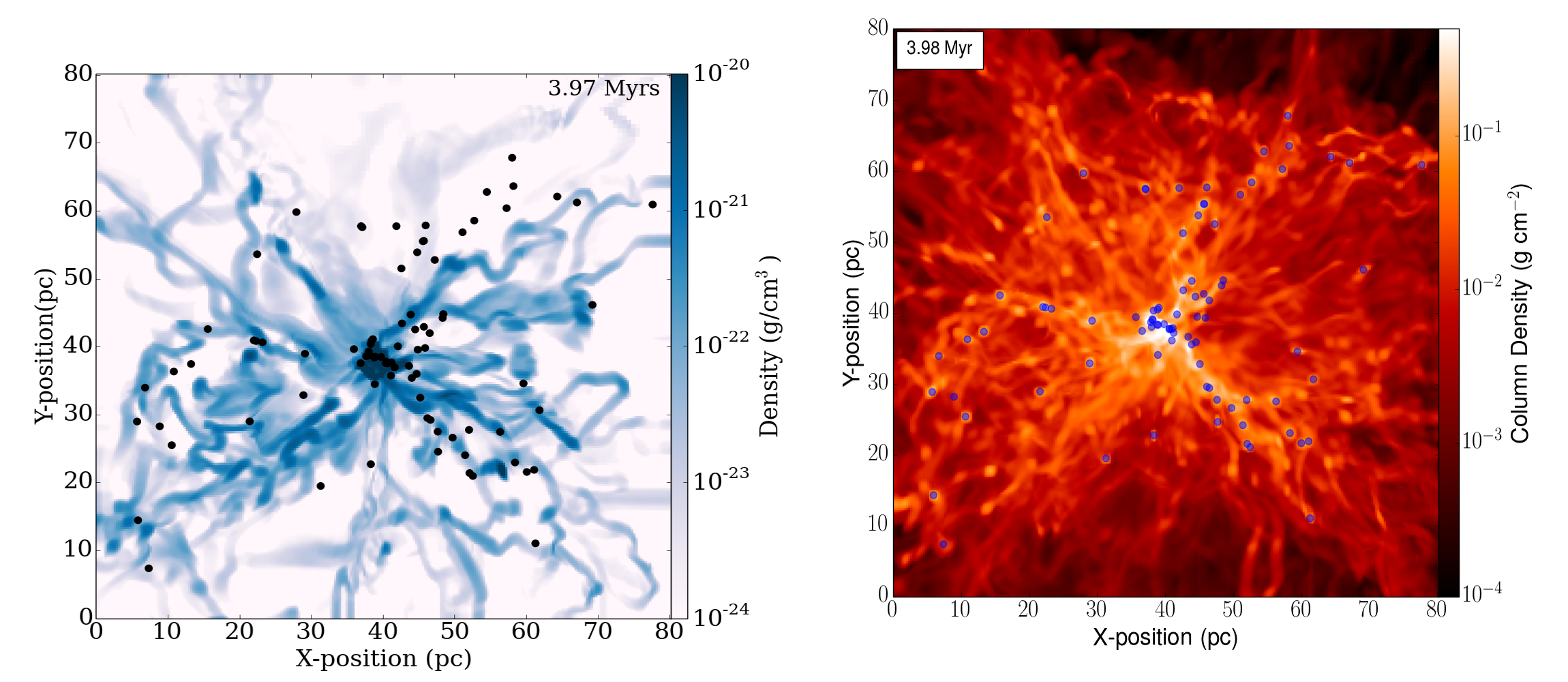} \\
\end{tabular}
\caption{Density slices through the centre of the z-axis (left) and column density projections (right) for the simulation with $\alpha_0$ $=$ 1. Cluster locations are projected onto the viewing plane and are plotted as black circles for the slice plots and blue circles for the column density plots.}
\label{fig:slice_coldens_alpha1}
\end{figure*}

\begin{figure*}
\begin{tabular}{c}
\includegraphics[width=0.8\linewidth]{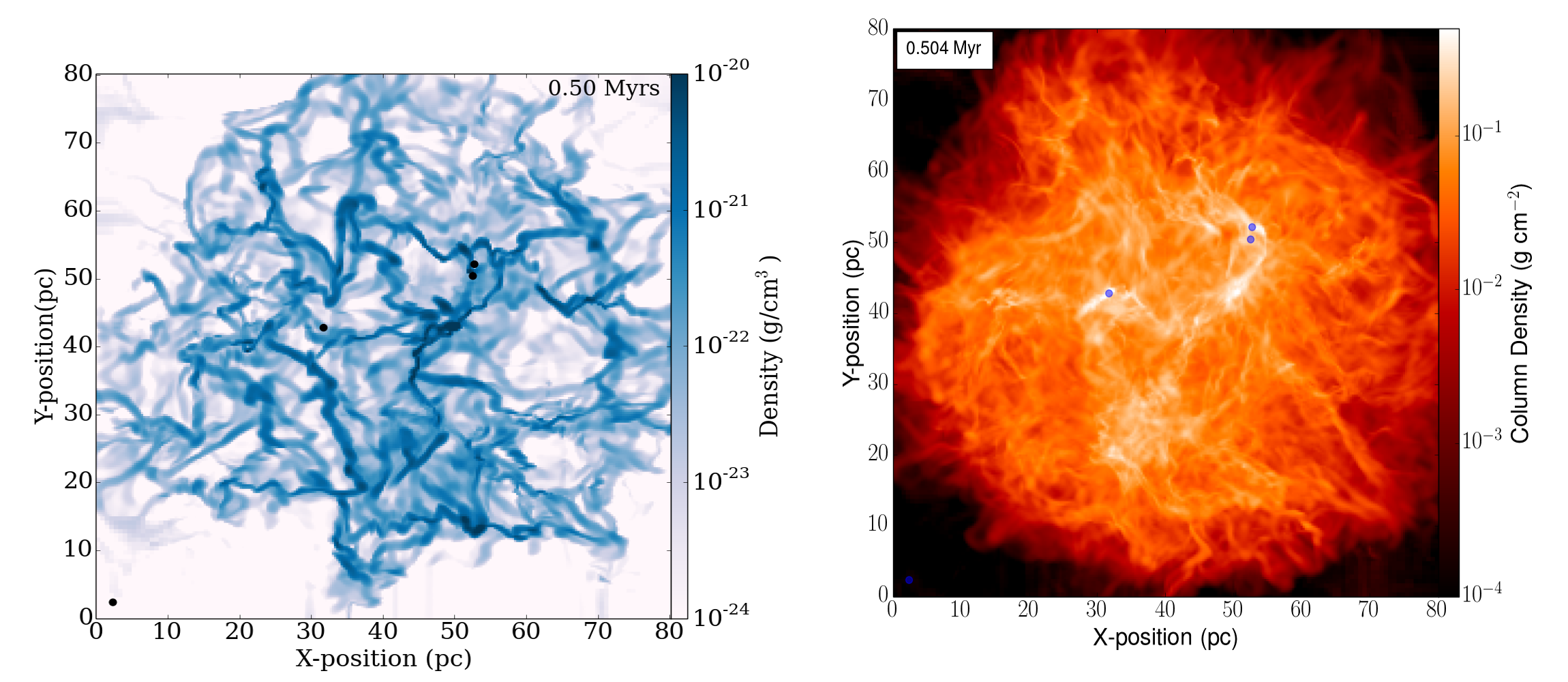} \\
\includegraphics[width=0.8\linewidth]{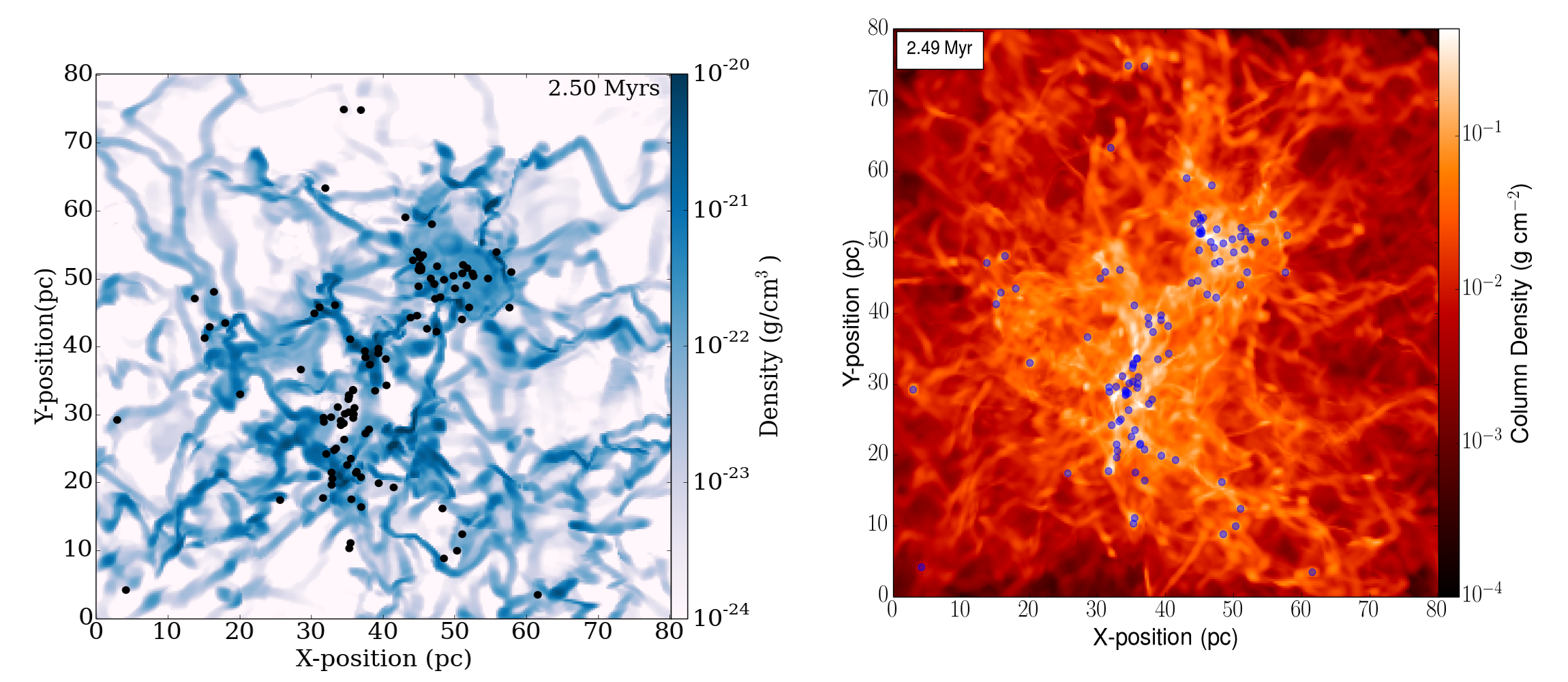} \\
\includegraphics[width=0.8\linewidth]{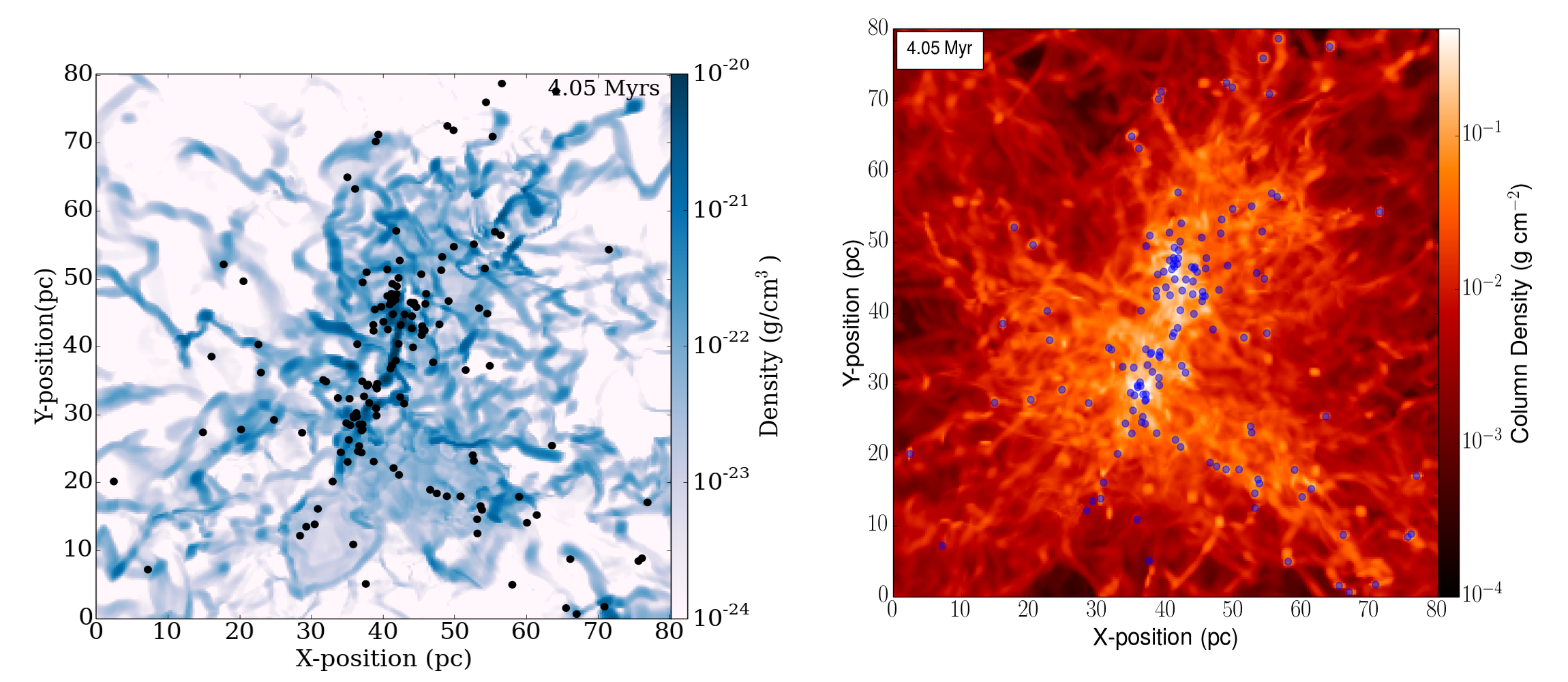} \\
\end{tabular}
\caption{Density slices through the centre of the z-axis (left) and column density projections (right) for the simulation with $\alpha_0$ $=$ 3. Cluster locations are projected onto the viewing plane and are plotted as black circles for the slice plots and blue circles for the column density plots.}
\label{fig:slice_coldens_alpha3}
\end{figure*}

\begin{figure*}
\begin{tabular}{c}
\includegraphics[width=0.8\linewidth]{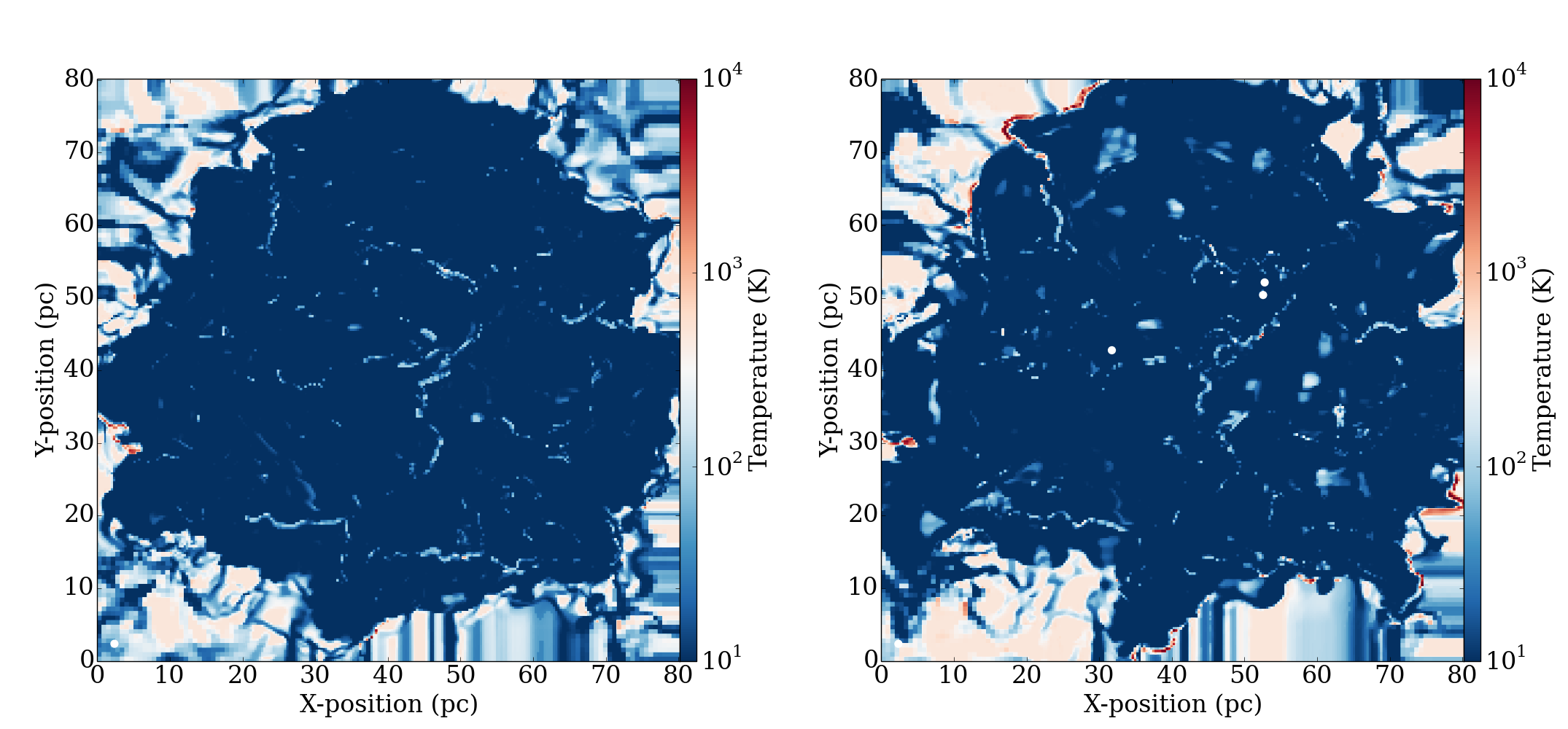} \\
\includegraphics[width=0.8\linewidth]{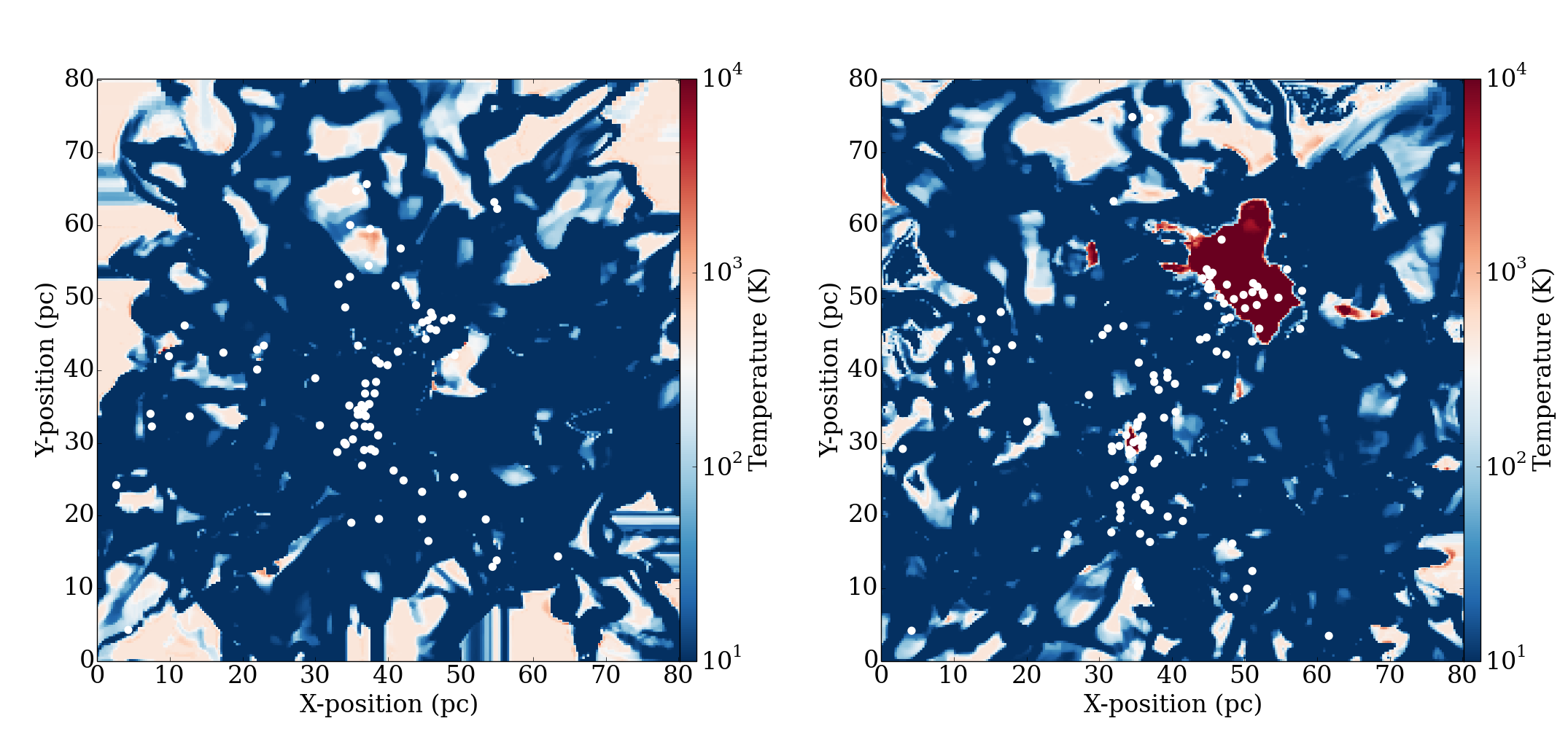} \\
\includegraphics[width=0.8\linewidth]{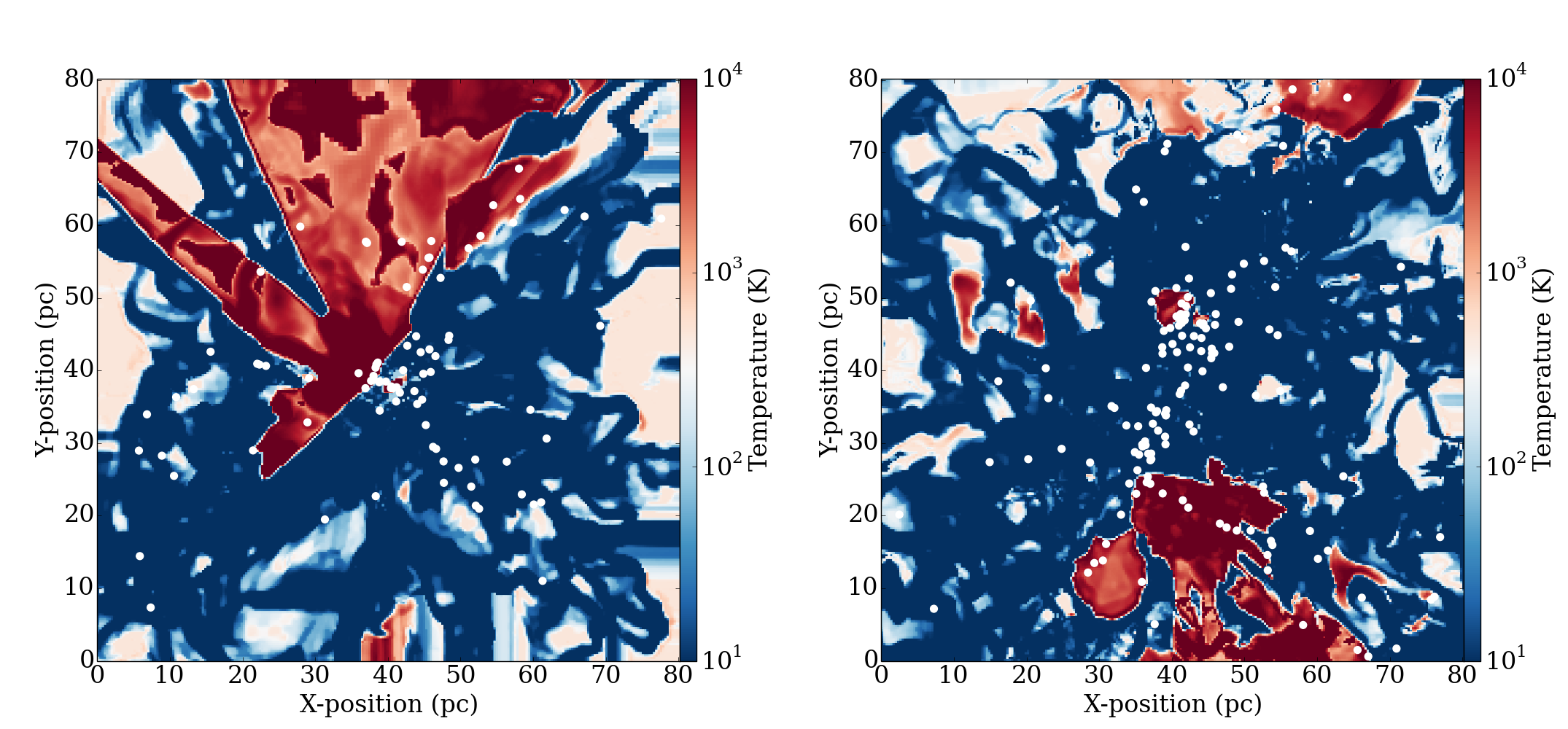} \\
\end{tabular}
\caption{The corresponding temperature slices to Figure \ref{fig:slice_coldens_alpha1} (left) and Figure \ref{fig:slice_coldens_alpha3} (right). Cluster particles are now shown in white for ease of viewing.}
\label{fig:slicetemp}
\end{figure*}

A three-dimensional view showing the evolution of the virialized simulation ($\alpha_0$ $=$ 1) at three different times can be seen in Figure \ref{fig:3d}. Three different density contours are plotted with 
densities ranging from 1.0$\times$10$^{-23}$ to 3.0$\times$10$^{-20}$ g$\cdot$cm$^{-3}$. These specific densities were chosen to highlight the sparse, low density 
gas, the intermediate density filaments, and the dense cores which form primarily in the center of the simulation. The bounding box has a side length of 80.2 pc and the 
location of the cluster particles are overplotted with black circles. The red regions in Figure \ref{fig:3d} represent gas which has an ionization fraction greater than 
95$\%$, highlighting the HII regions produced by the cluster particles. 

The first panel in Figure \ref{fig:3d} shows that the gas quickly breaks up into filaments. A total of 17 cluster particles have formed by this time but have not 
produced enough stars to produce HII regions. The second panel, plotted at 2.68 Myr, still shows a filamentary network of gas but one which is more centrally condensed compared
to the previous panel due to the global collapse of the gas. Dense cores are also now visible, scattered throughout the central region of the box where filaments intersect. A total
of 84 particles are present at this time, some of which have produced enough stars to begin feeding back on their environment to produce small, localized HII regions. The total number of cluster particles remains roughly the same with a total of 88 particles at 3.65 Myr, shown in the bottom panel of Figure \ref{fig:3d}, but the stellar content in these clusters has
grown significantly, allowing larger HII regions to form which cover a large fraction of the box. One region in particular has expanded outward from the central region into
low density gas and extends to the edge of the simulation volume.

To give a complementary view to Figure \ref{fig:3d}, we show density slices through the center of the z-axis and column density projections in Figure \ref{fig:slice_coldens_alpha1}, for $\alpha_0$ $=$ 1, and Figure \ref{fig:slice_coldens_alpha3}, for $\alpha_0$ $=$ 3.
 The corresponding temperature slices are in Figure \ref{fig:slicetemp}. We show an example of a bound and unbound ($\alpha_0$ $=$ 3) 
simulation illustrate the effects that initial boundedness plays on the evolution of the cloud. The locations of the cluster particles are projected onto the viewing plane in all images. 

Like the three-dimensional images discussed above, both clouds quickly break up into filaments by $\sim$0.5 Myr. Since the unbound cloud has a higher Mach number, the 
filaments are more pronounced with more low density gas filling the voids between the filaments. The temperature profiles look similar at this point, with cold 10 K gas in the center
surrounded by warmer $\sim$300 K gas. There are visible shocks, however, in only the unbound simulation, along the periphery of the cloud. Comparing the two simulations at the 
earliest times also shows that only the unbound cloud has formed cluster particles. This can again be attributed to its higher Mach number, leading to stronger density enhancements
which fragment into particles. 

The middle panels of Figure \ref{fig:slice_coldens_alpha1}, \ref{fig:slice_coldens_alpha3}, and \ref{fig:slicetemp}, plotted at $\sim$2.5 Myr, show that both simulations have produced many clusters, totalling
81 and 107 for the bound and unbound simulation, respectively. The clusters are grouped more heavily towards the centre of the volume for the bound run, indicative of collapsing
gas resulting in higher central densities and a more centrally condensed grouping of clusters. An HII region is also visible in the middle panel of Figure \ref{fig:slicetemp}. 
The corresponding density slice shows that the HII region has disrupted the filaments, effectively smearing them out to a more uniform medium with density enhancements around
the edges. 

The final panel, shown at $\sim$4 Myr, shows more marked differences between the initially bound and unbound clouds. The bound cloud is now even more centrally condensed due to global collapse, 
with one dense central core from which filaments radiate outwards. A large grouping of clusters remains in the center, allowing further accretion. The outskirts are devoid of 
dense gas, which has allowed a large HII region to extend northwards, away from the central condensation, where there is less gas to shield these regions from the ionizing photons
produced by massive stars in the central grouping. The unbound simulation, on the other hand, has dense gas pervading a large fraction of the box and lacks a dense, central 
core. The number of clusters has grown slightly for a total of 87 and 140 for the bound and unbound run, respectively. This suggests that the majority of 
cluster formation occurs during the early evolution of the clouds.   
       
\begin{figure}
 \begin{center}
  \includegraphics[width=0.9\linewidth]{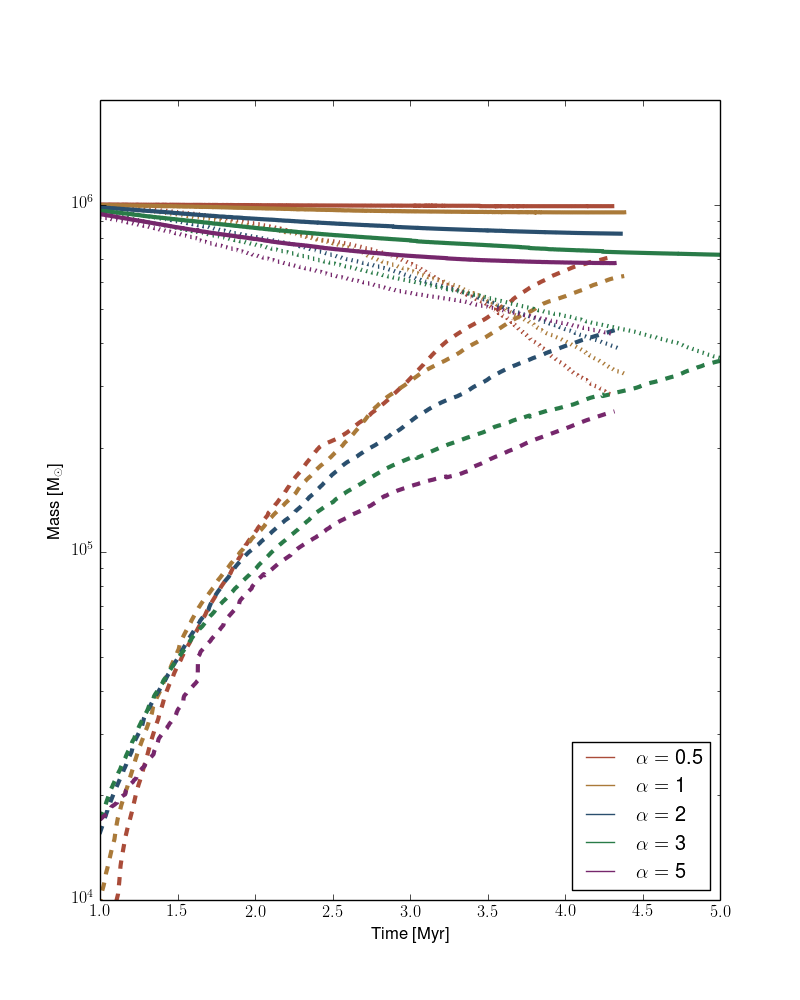}
 \end{center}
 \caption{Mass components in 5 simulations with different $\alpha_0$'s. The dashed lines show the total mass in cluster particles. The dotted lines show the total gas mass in the simulation volume. The total mass in the simulation is the sum of gas and particles and is shown by the solid lines. Note that mass is lost from the simulations, particularly in the unbound cases, because outflow conditions are present.}
\label{fig:mass_components}
\end{figure}

\begin{figure*}
 \begin{center}
  \includegraphics[width=0.65\linewidth]{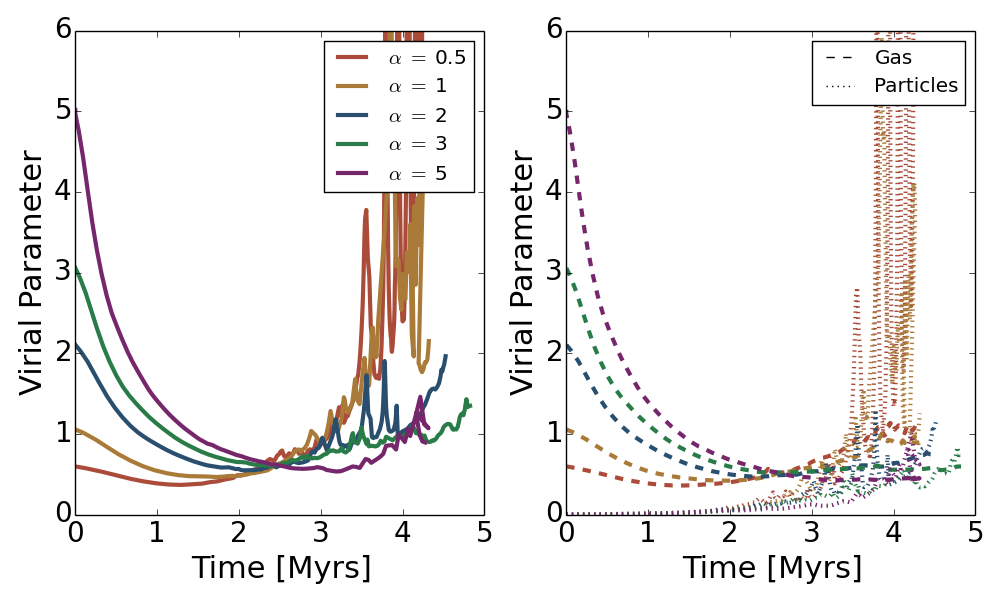}
 \end{center}
 \caption{Left: Evolution of the virial parameter over time including both gas and particles. Right: Individual contributions to $\alpha$ from the gas (dashed) and the cluster particles (dotted). See text for more detail on how the virial parameter was split.}
 \label{fig:alpha}
\end{figure*}

The above visualizations provide a qualitative description of the evolution of a small subset of the completed simulations. A quantitative comparison showing both the gas and cluster particle evolution for
all simulated clouds, with radiative feedback included, can be seen in Figure \ref{fig:mass_components}. The total mass in cluster particles is shown by the dashed lines, the total gas 
mass present in the simulation volume is shown by the dotted lines, and their sum is the shown by the solid lines. Since outflow boundary conditions are present, the solid 
lines do not remain constant at 10$^6$ M$_{\odot}$ because gas can leave the simulation volume. The total amount of mass lost from the simulation depends heavily on $\alpha_0$. From Figure \ref{fig:mass_components}, it can be seen that the initially bound runs lose a negligible amount of mass. The unbound runs, on the other hand, 
lose a significant amount. For example, the case with an initial virial parameter of 5 loses $\sim$30$\%$ of the total mass by the end of the simulation.

Clearly $\alpha_0$ plays an important role in the long term evolution of the cloud. However, the virial parameter does not remain constant over the course of the simulation because the initial supersonic turbulence is damped via shocks. As the turbulence is damped, the velocity dispersion of the gas decreases which results in a decreasing $\alpha$. The loss of high velocity gas from the simulation volume and the formation of cluster particles further modifies $\alpha$.

In Figure \ref{fig:alpha}, we plot the global virial parameter as a function of time on the left, and the contributions to $\alpha$ by the gas and cluster particles on the right. The global virial parameter, and the contributions from particles and gas, are expressed via,

\begin{equation}
\alpha = 2\frac{E_{kin}}{|E_{grav}|} = 2\frac{E_{kin,p}}{|E_{grav}|} + 2\frac{E_{kin,g}}{|E_{grav}|} = \alpha_{p} + \alpha_{g}
\end{equation}

\noindent where E$_{kin}$ is the kinetic energy, E$_{grav}$ is the gravitational potential energy, and the subscripts $g$ and $p$ refer to gas and particles.  

We see from Figure \ref{fig:alpha} that $\alpha$ initially decreases as the turbulence damps. The rate at which $\alpha$ decreases is greater for unbound simulations. All simulations tend towards a virial state (ie. $\alpha$ $=$ 1), and cross at approximately 2.3 Myr. The tendency towards a virial state is likely due to gravitational collapse after the decay of the initial turbulence which, in turn, drives enough internal turbulence to maintain $\alpha$ at 1 \citep{Ballesteros,Heitsch2013}. Because the virial parameters intersect at similar times, the simulations with higher $\alpha_0$ decrease more rapidly compared to lower $\alpha_0$ simulations. This is attributed to the turbulent decay time, $\tau_{dec}$, which is given by $\tau_{dec}$ $\sim$ $L$/($M c_s$) \citep{Tilley2004}, where $L$ is the box size, $M$ is the turbulent Mach number, and $c_s$ is the sound speed. Measuring the turbulent decay times from Figure \ref{fig:alpha} shows a relation that is roughly inversely proportional to the initial Mach 
number, as suggested by $\tau_{dec}$. The relation is not exact, however, due to varying degrees of mass loss from bound and unbound models.

The right hand panel shows that this early phase, characterized by a decreasing $\alpha$, is dominated almost exclusively by the gas energetics. A significant number of cluster particles have formed by this time (eg. $\sim$10$\%$ of all mass is in the form of cluster particles by 2.3 Myr for the simulation with $\alpha_0$ $=$ 0.5), but they contribute negligibly to the overall $\alpha$. After approximately 3 Myr, however, particle dynamics begin to play a significant role in the energetics, leading to a rise in $\alpha$. The cluster particle contribution even outweighs the gaseous contribution for the initially bound runs. 

The role of the particle contributions to the global energetics points towards interesting dynamical differences between the simulations. We find that, in the bound simulations, the cluster particles are more centrally condensed relative to unbound simulations. This tighter grouping leads to stronger two-body interactions which result in velocities in excess of 50 km/s. These large particle velocities are responsible for the dominant contribution to $\alpha$. For comparison, the velocities in unbound simulations do not exceed 20 km/s. We leave the discussion of particle dynamics, and its connection to $\alpha_0$, to a later paper. 

\begin{figure*}
 \begin{center}
  \includegraphics[width=0.65\linewidth]{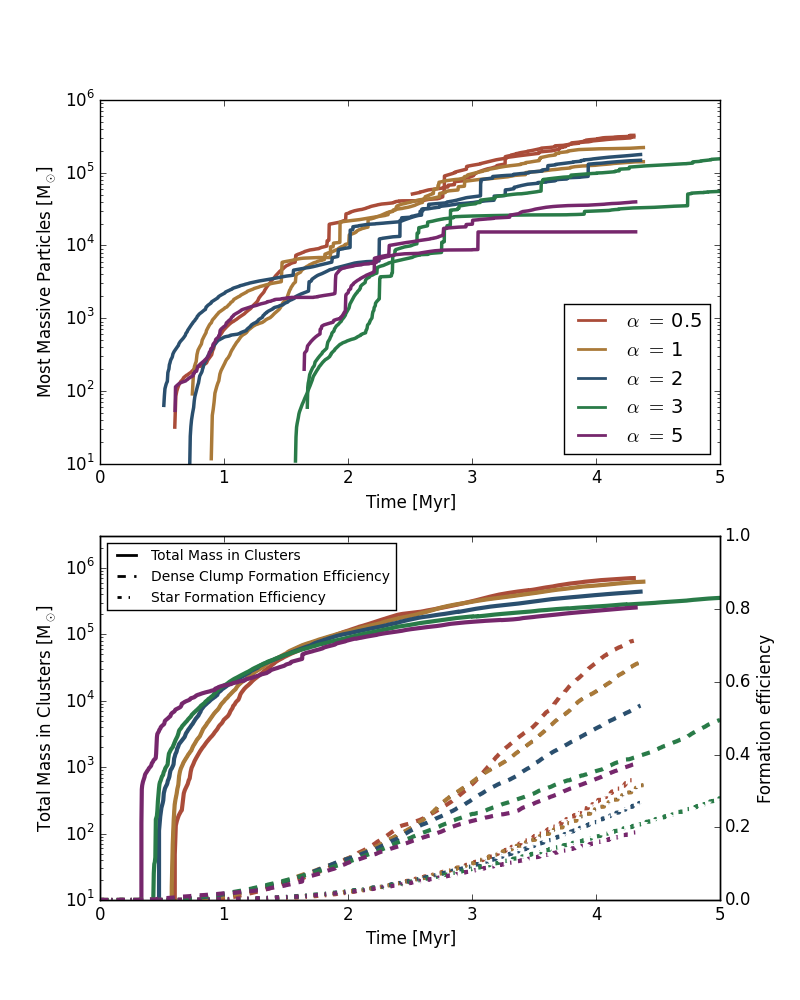}
 \end{center}
 \caption{Top: The evolution of the total mass for the two most massive cluster particles that survive until the end of the simulation. Discrete jumps in total mass are due to cluster particle merging. Bottom: Solid lines represent the total mass contained in cluster particles and corresponds to the left axis. Dashed lines represent the dense clump formation efficiency, $\epsilon_{cl}$. Dashed-dotted lines represents the star formation efficiency, $\epsilon_{sf}$.}
\label{fig:boundedness}
\end{figure*}

\subsection{Formation Efficiencies and Particle Evolution}

The discussion so far has been focused on global properties, such as the total mass in cluster particles, rather than the properties of individual objects. We move now to a discussion of the cluster particles themselves. The top panel of Figure \ref{fig:boundedness} shows the time evolution of the two most massive cluster particles that survive to
the end of the simulation for all clouds. Since the particles are not allowed to lose mass, these curves are monotonically increasing
as a function of time. All curves also show a series of discrete jumps in mass. These jumps are due to merging events, as described in Section \ref{methods}. It can be concluded from these curves that \textit{merging plays a significant role in determining the mass of the most massive particles in these simulations}. This behaviour has also been observed in cluster-scale simulations \citep{Maschberger10}.

An interesting trend is noted in the top panel of Figure \ref{fig:boundedness}, namely that the simulations with a lower $\alpha_0$ tend to produce more massive cluster particles. The maximum mass cluster particles, including the mass of gas and stars, range from 3.26$\times$10$^4$ M$_{\odot}$ to 3.27$\times$10$^5$ M$_{\odot}$ for simulations with an $\alpha_0$ of 5 and 0.5, respectively. This trend can be attributed to the central concentration of dense gas that forms as a result of gravitational collapse in the bound simulations, allowing the most massive particles to continue accreting strongly from their surroundings. This suggests that the distribution of cluster particle masses depends on $\alpha_0$, which is discussed further below. 

The solid lines in the bottom panel of Figure \ref{fig:boundedness} show the total mass of all cluster particles 
(shown on the left y-axis). There is a spread between simulations in both the total mass in cluster particles, and the time at which cluster formation begins. There is a spread of $\sim$260 kyr between when the most unbound cloud ($\alpha_0$ $=$ 5) and the most bound cloud ($\alpha_0$ $=$ 0.5) begin to form particles. 
  
While the unbound simulations form the first cluster particles, the bound simulations have more mass contained in particles overall. This can be seen by the dashed lines in the
bottom panel of \ref{fig:boundedness}. We refer to these lines as the "dense clump formation efficiency", hereafter $\epsilon_{cl}$, and are calculated by dividing the total mass in cluster particles over the initial
mass of the cloud, ie.

\begin{equation}
\epsilon_{cl} = \frac{\sum_{i} M_i}{M_{total}} =  \frac{\sum_{i} M_i}{10^6 M_{\odot}}
\end{equation}

\noindent where $M_i$ is the mass of the $ith$ particle and the total cloud mass of 10$^6$ M$_{\odot}$ has been shown for clarity. We have chosen the nomenclature of "dense clump formation efficiency", rather than a cluster formation efficiency, because the mass contained in these particles is not solely in stars, but also a reservoir of unused gas which may be used for future star formation.

Since $\epsilon_{cl}$ closely follows the total mass contained in particles, we see a similar spread between simulations. The early evolution is similar, until
about 2 Myr, when the dense clump formation efficiencies begin to diverge. The final efficiencies cover a range of 37$\%$, for an $\alpha_0$ of 5, to 71$\%$, for an $\alpha_0$ of 0.5 with the unbound runs ending up less efficient than the bound runs. This is consistent with earlier simulations which demonstrated that particle formation efficiencies, over a fixed time, are significantly reduced as the Mach number is increased \citep{KHM2000}.

\begin{figure}
 \begin{center}
  \includegraphics[width=0.9\linewidth]{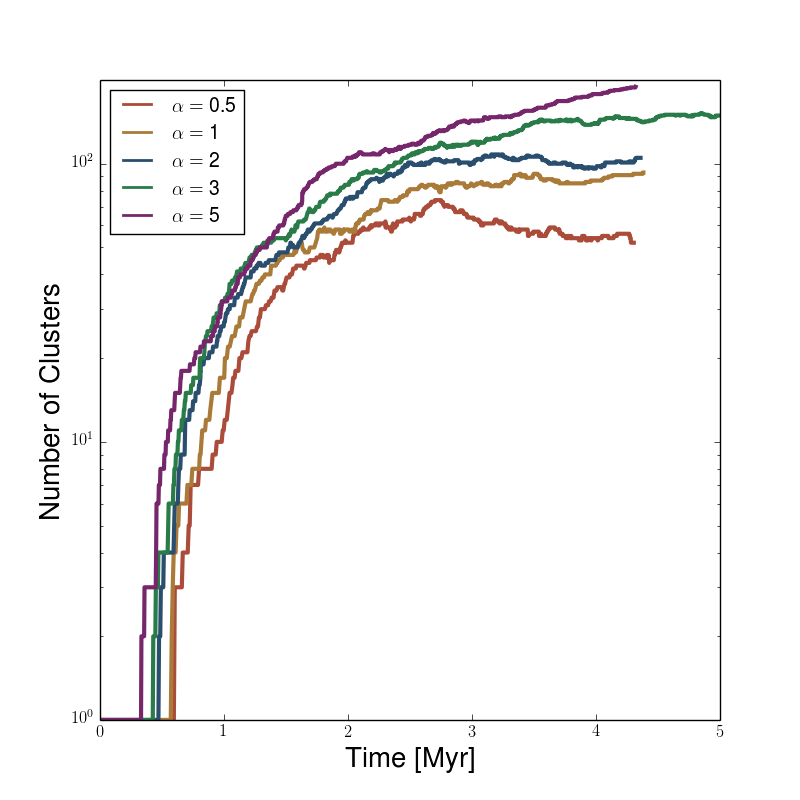}
 \end{center}
 \caption{The total number of cluster particles as a function of time. The total number can increase via forming new particles, and decrease via mergers or particles leaving the simulation volume.}
 \label{fig:nstars}
\end{figure}

Our $\epsilon_{cl}$ values can be compared to the observationally measured "dense gas mass fractions" (DGMFs) in nearby molecular clouds.
Recent work by \cite{Abreu} has measured the fraction of dense gas, defined to be gas above an extinction of $A_{V}$ $>$ 7 corresponding to gas at 
densities of $n$ $>$ 10$^4$ cm$^{-3}$, in local molecular clouds with masses ranging from 200 to 2$\times$10$^5$ M$_{\odot}$. We note that our simulations 
have greater masses than the clouds presented in \cite{Abreu}, but the observations still provide a useful point of comparison. The observed clouds span a range of 
evolutionary stages from starless clouds to star forming clouds with well developed HII regions. They find that the DGMF covers a large range, with a maximum value of 
$\sim$0.8. The average DGMF over the entire mass range is found to be 0.39 with a cloud-to-cloud scatter of 0.28, and the higher mass clouds have DGMFs close to the maximum value. We see that our own 
DGMFs, as measured by the total mass in cluster particles, fall within the observed range. The simulation having the closest DGMF to the observational average is the unbound simulation $\alpha_0$ $=$ 3. This suggests that initially unbound molecular clouds more closely reproduce the properties of locally observed clouds.

This point can be elaborated further by examining the dash-dotted lines in Figure \ref{fig:boundedness}. These lines represent the star formation efficiency of the clouds ($\epsilon_{sf}$), defined as the total mass in stars within cluster particles divided by the initial cloud mass. These curves show similar trends to $\epsilon_{cl}$ 
in the sense that higher initial virial parameters correspond to lower efficiencies. The curves take longer to diverge, however, as a consequence of our subgrid model 
which converts the cluster mass to stars gradually over time. The final $\epsilon_{sf}$ ranges from 18\% to 34\%. These values are higher 
than the global $\epsilon_{sf}$ in observed molecular clouds which typically span a range of 1-5$\%$ (eg. \cite{Duerr}). Since the highest virial parameter produces the lowest 
efficiency which is close to the observed values, this again suggests that a high initial virial parameter, in combination with radiative feedback, is required to match 
observations. There are, however, other forms of feedback that this work neglects which could also play a role in reproducing observed $\epsilon_{sf}$ values.  

When discussing cluster particles properties up to this point, we have focused solely on mass. To fully understand the resulting mass distributions, however, the total number of cluster sink particles needs to be discussed. In Figure \ref{fig:nstars}, we show the total number of cluster particles as a function of time. The number of particles can either increase via formation, or decrease via mergers or particles leaving the simulation volume. It can be seen that a large burst of cluster formation occurs early but levels out to roughly constant values around 2 Myr. The final number of clusters ranges from 52 for $\alpha_0$ $=$ 0.5 to 189 for $\alpha_0$ $=$ 5. As noted earlier, we see that initially unbound simulations begin to form cluster particles earlier. These curves do not cross, however, which is in contrast to the mass curves discussed in Figure \ref{fig:boundedness}. This means that while the bound simulations contain the most mass in cluster particles, they have the lowest number of particles 
overall. The bound simulations are therefore expected to have more massive particles, on average, than unbound simulations. This has will be discussed further in the following section.

\subsection{Radiative Feedback Effects}

\begin{figure*}
 \begin{center}
  \includegraphics[width=0.65\linewidth]{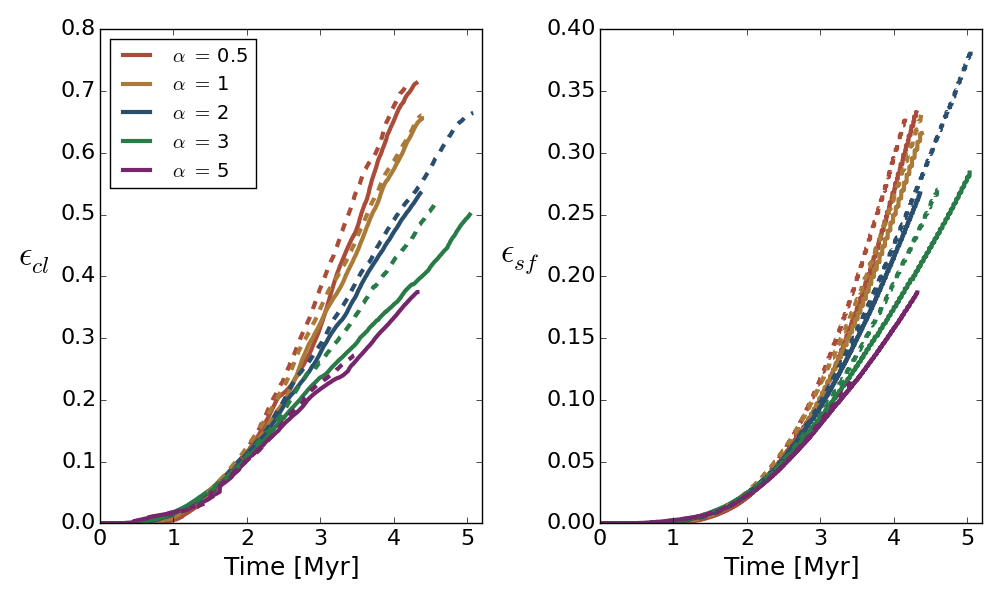}
 \end{center}
 \caption{The dense clump formation efficiency (left) and the star formation efficiency (right) for simulations with radiative feedback included, shown by solid lines, or not included, shown by the dashed lines.}
 \label{fig:rad_vs_hydro}
\end{figure*}

The above discussion has been focused on the role that $\alpha_0$ plays in shaping cloud properties and the resulting efficiencies. We now move on to discuss the impact of radiative feedback on these properties. To do this, we have completed an additional five simulations which are identical to those presented above but with radiative feedback not included. We refer to the simulations with radiative feedback included as "RHD" (Radiation Hydrodynamics) and the simulations with radiative feedback turned off as "HD" (Hydrodynamics).

The main effect of radiative feedback is the suppression of star, or cluster, formation via the heating and ionization of the gas surrounding the newly formed stars. The strength of this suppression, however, is not fully understood and depends on the environment being modelled. For example, \cite{Dale2005} found that ionization feedback had a negligible effect on global SFEs in initially gravitationally bound clouds. In contrast, later work by \cite{2012MNRAS.427.2852D} showed that ionization feedback can reduce the SFE by up to 50\% in smaller, 10$^4$ M$_{\odot}$ clouds. Since it is clear that the GMC environment is a crucial component in determining the role of radiative feedback, we examine whether this role differs in 10$^6$ M$_{\odot}$ GMCs with various $\alpha_0$. We have also included a rudimentary treatment of radiation pressure in our radiative transfer scheme which was not included in these previous works.

\begin{figure}
 \begin{center}
  \includegraphics[width=0.9\linewidth]{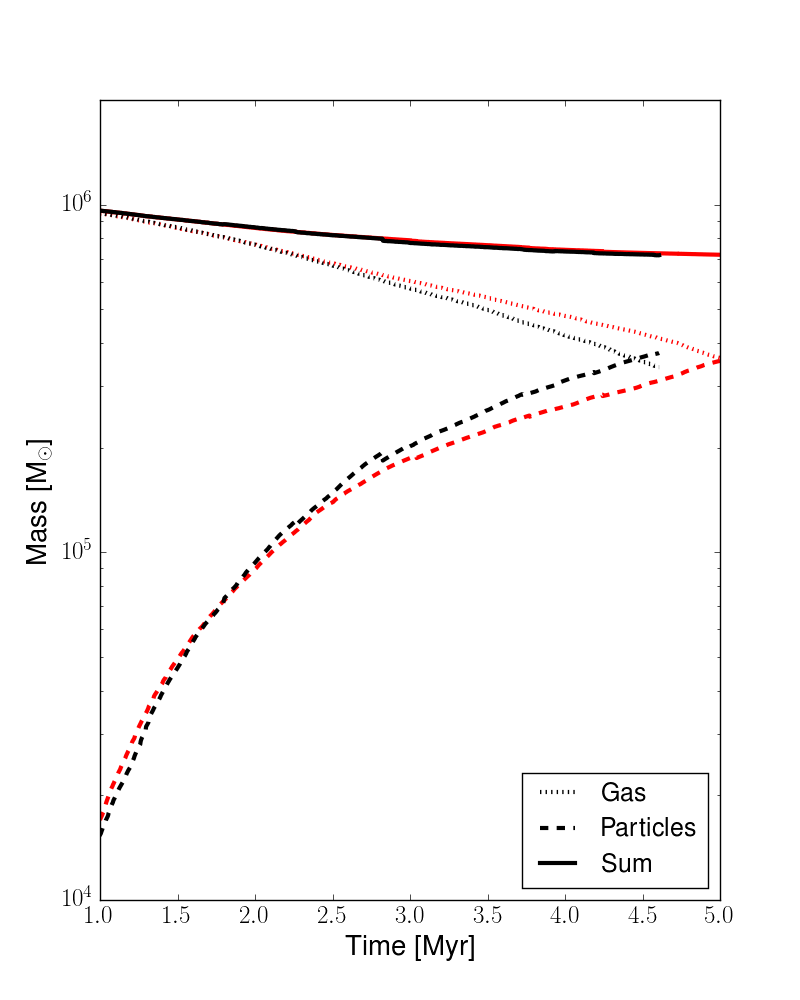}
 \end{center}
 \caption{Mass components in the $\alpha_0$ $=$ 3 simulations (similar to Figure \ref{fig:mass_components}). Red lines represent the RHD simulations including radiative feedback, while the black lines represent the HD simulations not including radiative feedback.}
 \label{fig:mc_alpha3}
\end{figure}

We compare the dense clump formation efficiencies, $\epsilon_{cl}$, and star formation efficiencies, $\epsilon_{sf}$, between RHD and HD simulations in Figure \ref{fig:rad_vs_hydro}. Solid lines represent the RHD simulations and dashed lines represent the HD simulations. We see that the inclusion of radiative feedback plays a small role in determining $\epsilon_{cl}$ and $\epsilon_{sf}$ in the early evolution of a GMC, regardless of initial boundedness. The largest difference in $\epsilon_{cl}$ occurs in the simulation with an $\alpha_0$ of 3, with a difference of $\sim$10\% between the RHD and HD simulation. The other simulations show that radiative feedback does suppress $\epsilon_{cl}$ but only by $<$3\%. Since $\epsilon_{sf}$ closely mirrors the evolution of $\epsilon_{cl}$ but with a time delay, the difference in SFEs between RHD and HD simulations is less noticeable. 

Overall, these results suggest that radiative feedback plays a minimal role in the early evolution of a massive GMC. Comparing these results to those presented in Figure \ref{fig:boundedness}, it is clear that $\alpha_0$ is the major contributing factor to the final efficiencies. Visual examination of Figure \ref{fig:slicetemp}, however, shows that radiative feedback is producing large scale HII regions which are evidently not strongly suppressing the global efficiencies in the cloud. The filamentary and porous nature of the cloud is likely limiting the role of radiative feedback.

\begin{figure*}
 \begin{center}
  \includegraphics[width=0.65\linewidth]{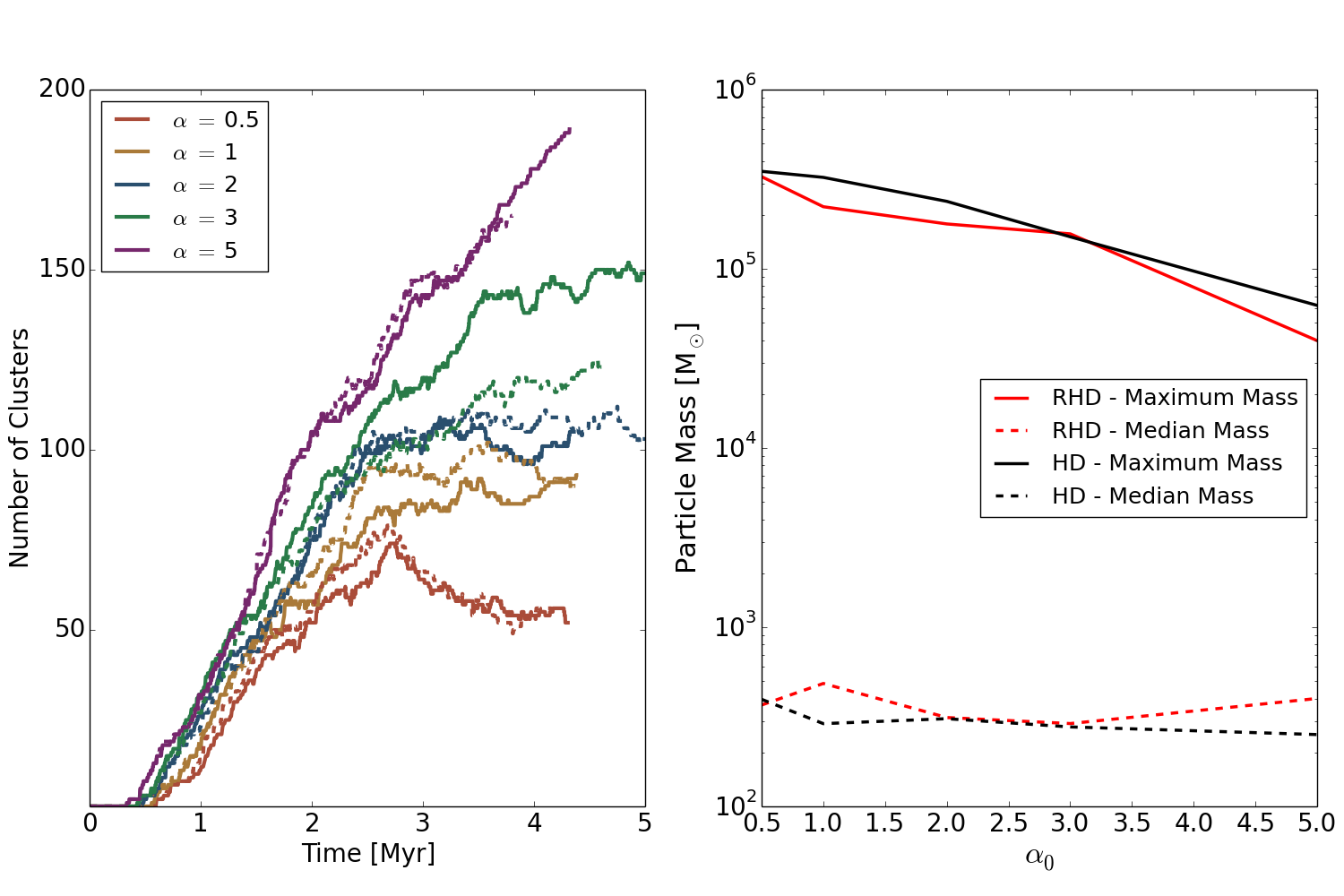}
 \end{center}
 \caption{Left: The number of cluster particles as a function of time for RHD simulations, shown by the solid lines, and HD simulations, shown by the dashed lines. Right: The maximum and median mass cluster particles for simulations with different $\alpha_0$ values.}
 \label{fig:clusters_hyd}
\end{figure*}

It is possible that if we followed the evolution of the cloud over longer timescales, we would see a larger suppression in efficiencies due to the lower overall density of the cloud which can be more easily disrupted by radiative feedback. Since we did not follow the cloud evolution to $>$5 Myr, at which point Supernovae are expected to begin injecting significant amounts of energy and momentum into the surrounding gas \citep{Starburst99}, we can only conclude that radiative feedback does not greatly suppress cluster and star formation in the early evolution of a GMC.         

To show in which ways radiative feedback is suppressing the formation of particles, we focus on the simulation which had the largest difference between the RHD and HD simulations, namely the simulation with an $\alpha_0$ of 3. The observed difference in efficiency could either be due to radiation unbinding gas which then leaves the simulation volume, or through the heating and ionization of gas locally which prevents future particle formation and limits the accretion rate onto existing particles. To see which of these scenarios is more likely, we examine how the mass is divided among gas and particles in Figure \ref{fig:mc_alpha3}.

Figure \ref{fig:mc_alpha3} shows that the RHD simulation has comparably less mass contained in particles and more gas mass than the HD simulation. However, the sums of these components, shown by the solid lines, are nearly identical, indicating that the RHD and HD runs have lost the same amount of mass from the simulation volume. From this we can conclude that radiative feedback is not driving outflows from the GMC, but is instead suppressing the formation and accretion of cluster particles. We note that this may be a product of our large cloud masses (10$^6$ M$_{\odot}$) which have correspondingly deep potential wells which can trap the gas even under the influence of radiative feedback. Outflows may be more relevant in the early evolution of less massive GMCs where the energy and momentum input from stars may be sufficient to unbind the cloud.

While global efficiencies are not greatly influenced by the inclusion of radiative feedback, it is possible that the cluster properties formed in an RHD versus an HD simulation may differ. In Figure \ref{fig:clusters_hyd}, we plot cluster properties for HD and RHD simulations to examine differences that arise. On the left, we show the total number of particles in all simulations. On the right, we plot the maximum and median mass cluster particles as a function of $\alpha_0$. 

The left hand panel of Figure \ref{fig:clusters_hyd} shows no clear trend in the number of cluster particles in RHD versus HD simulations. The simulations with an $\alpha_0$ of 0.5 and 5 show no difference between the number of particles with radiative feedback included. The simulations with an $\alpha_0$ of 1 and 2 show an excess of clusters in the HD simulations at intermediate times but the difference becomes negligible at late times. Only the simulations with an $\alpha_0$ of 3 have different numbers of clusters at late times, with the RHD simulation forming $\sim$30 more cluster particles than the HD simulation. 

The right hand panel of Figure \ref{fig:clusters_hyd} shows the maximum and median mass cluster particles versus $\alpha_0$ in both HD (black) and RHD (red) simulations. The maximum mass curves (solid lines) indicate that, except for the $\alpha_0$ $=$ 3, HD simulations tend to produce more massive cluster particles than their RHD counterparts. This effect has been observed in other simulations of radiative feedback in which the distribution of particle masses is shifted to lower values \citep{2012MNRAS.427.2852D}. The median mass clusters show the opposite trend, with the RHD median masses being greater than the HD median masses. This indicates that the mass distribution is not simply being shifted to lower masses since the same trend would be seen in the median mass clusters as in the maximum mass clusters. It should be noted that since our subgrid model only includes radiative feedback from 
massive clusters (M$_*$ $>$ 1000 M$_{\odot}$), the median mass clusters in the RHD simulations are not outputting radiation which may account for the observed trend reversal.

\subsection{Cluster Formation Thresholds} \label{thresholds}

\begin{figure*}
 \begin{center}
  \includegraphics[width=0.65\linewidth]{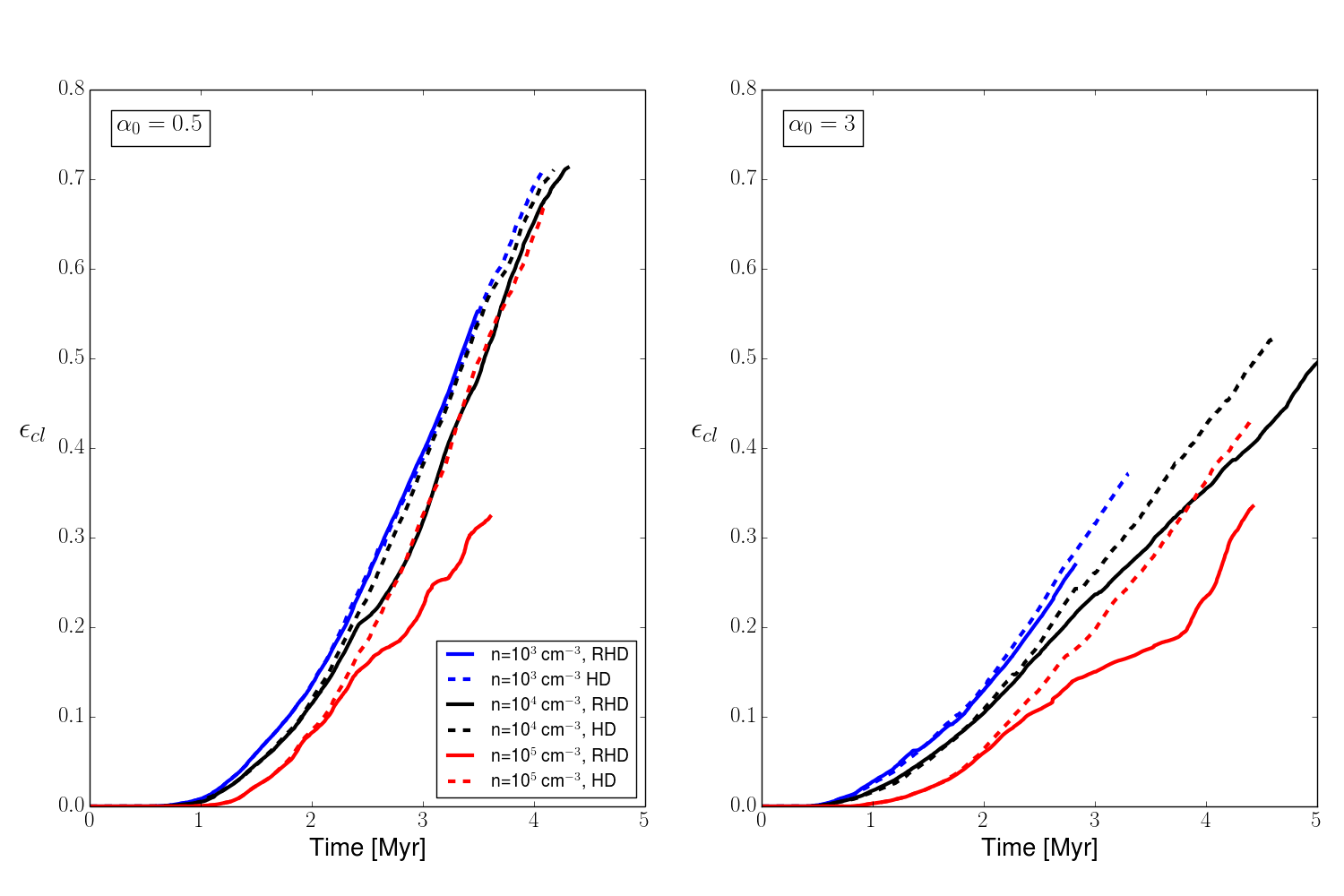}
 \end{center}
 \caption{The effects of radiative feedback on $\epsilon_{cl}$ from varying the particle formation threshold in a bound ($\alpha_0$ $=$ 0.5, left) and an unbound simulation ($\alpha_0$ $=$ 3, right). As described in Section \ref{methods}, the fiducial value we have used throughout is 10$^4$ cm$^{-3}$.}
 \label{fig:eff_nthresh}
\end{figure*}

As discussed above, recent studies suggest that the strength of radiative feedback in different GMC environments may vary. We have shown that this is not the case for 10$^6$ M$_{\odot}$ clouds which have different initial virial parameters. Another parameter, however, which also affects a cluster particle's local environment is the threshold density for formation. We have chosen a fiducial value of 10$^4$ cm$^{-3}$ (see Section \ref{methods}) motivated by observations of cluster forming regions. If, for example, a higher threshold density were chosen, the global GMC environment would be denser at the time of cluster formation and the radiation released by these clusters would be propagating into denser gas. This may have an important impact on the strength of radiative feedback. To examine the role that radiative feedback plays in clouds with different formation thresholds, we have resimulated a bound ($\alpha_0$ $=$ 0.5) and an unbound ($\alpha_0$ $=$ 3) cloud with formation thresholds of 10$^3$ and 10$^5$ 
cm$^{-3}$. The results of these simulations are presented in Figure \ref{fig:eff_nthresh}.

The onset of cluster formation differs among simulations with different formation thresholds. As expected, simulations with higher thresholds begin to form clusters later since the cloud has to collapse to higher densities before particles form. Comparing HD simulations, shown by dashed lines, and RHD simulations, shown by solid lines, we see that lower formation thresholds mirror the results presented in Figure \ref{fig:rad_vs_hydro} in the sense that radiative feedback plays a minor role in controlling $\epsilon_{cl}$. In fact, the differences in efficiency between RHD and HD simulations are reduced even further in the case of a low formation threshold. 

For the case of a high formation threshold, we see that the strength of radiative feedback is enhanced. The final efficiencies between RHD and HD simulation differ by 21\% and 10\% compared to 2\% and 8\% at the same time for the fiducial formation threshold of 10$^4$ cm$^{-3}$. These comparisons were at the time corresponding to the end of the shortest running simulation. The higher density immediately surrounding particles efficiently couples the radiation to the gas rather than the radiation streaming to low density voids perpendicular to filaments. The impact of radiative feedback on filaments is significantly more pronounced. This clearly indicates that the strength of radiative feedback is not a constant, but instead varies depending on the GMC environment. This may have important consequences for star formation in clumps and cores which are the densest regions in a molecular cloud. Indeed, observations of cores show that the densest cores ($n$ $\sim$ 10$^6$ cm$^{-3}$) have lifetimes of approximately a 
freefall time while less dense cores ($n$ $\sim$ 10$^{3-4}$ cm$^{-3}$) have lifetimes that are roughly ten times as long \citep{PPVI}. This may be due, in part, to the increased strength of radiative feedback in denser regions.

\begin{figure}
 \begin{center}
  \includegraphics[width=0.95\linewidth]{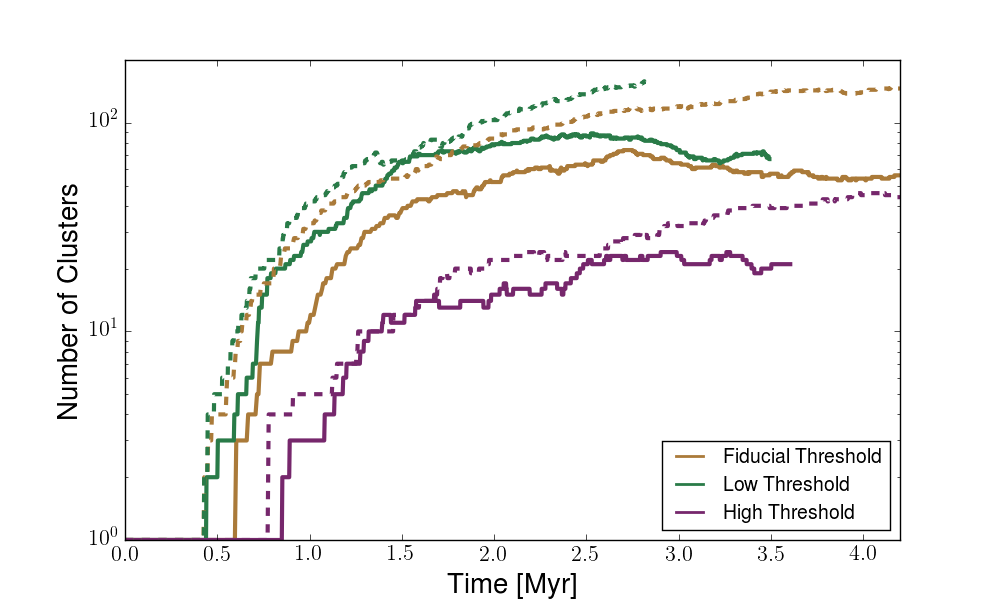}
 \end{center}
 \caption{The number of cluster particles versus time for different formation thresholds. Simulations with an $\alpha_0$ of 0.5 and 3 are shown by the solid and dashed lines, respectively.}
 \label{fig:nstars_nthresh}
\end{figure}

The most significant difference between simulations with different particle thresholds is the number of cluster particles that form, as shown in Figure \ref{fig:nstars_nthresh}. As the particle formation threshold increases, the number of particles decreases. For example, changing the formation threshold from 10$^3$ to 10$^5$ cm$^{-3}$ in the simulation with an $\alpha_0$ of 3 reduces the number of particles from 167 to 26. This trend is expected since the regions which form particles need to reach higher densities in the case of a higher threshold, and these regions become increasingly rare as the threshold increases.

While the particle formation efficiency is reduced in the case of a high formation threshold, it is not reduced by the factor of $\sim$6 that is seen in the particle number (for $\alpha_0$ $=$ 3). This is because the particles which form in the high threshold cases are, on average, more massive. Since a particle has a fixed size, and the regions out of which the particles form have higher densities for higher threshold cases, this result is expected.

The combination of fewer particles which are, on average, more massive for higher threshold cases could have a significant impact on the cluster particle mass distribution. This is discussed briefly in the next Section. In the remainder of this Section, however, we will continue to focus on simulations with the fiducial value of the formation threshold because it is observationally motivated, and a full exploration of the formation threshold parameter space is beyond the scope of this paper. 

\subsection{Cluster Properties}

\begin{figure}
 \begin{center}
  \includegraphics[width=0.95\linewidth]{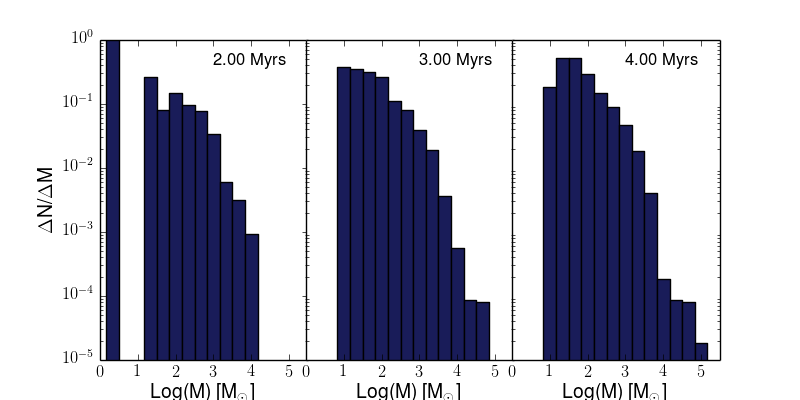}
 \end{center}
 \caption{Representative snapshots of the cluster mass function at various times (2, 3, and 4 Myr from left to right, respectively) for the simulation with an initial virial parameter of 3.}
 \label{fig:ECMF_evolution}
\end{figure}

We now move to discuss cluster properties in more detail with a focus on mass distributions and stellar content.

In Figure \ref{fig:ECMF_evolution}, we show representative snapshots of the normalized cluster particle mass function at 2, 3, and 4 Myr for the simulation with $\alpha_0$ $=$ 3. The other simulations show similar evolution, and a comparison between simulations is made below. The early evolution is dominated by low mass particles with the maximum particle mass reaching 10$^4$ M$_{\odot}$ by 2 Myr. While massive particles are present, the majority have masses on the order of Solar masses. 

As time progresses, as shown in the middle panel of Figure \ref{fig:ECMF_evolution}, the distribution shifts to higher masses because of both ongoing gas accretion and particle mergers. 

This same trend is evident in moving from 3 to 4 Myr. The highest mass particles have grown to approximately 10$^5$ M$_{\odot}$. The lowest bins have been depopulated, resulting in a turnover in the distribution just below 100 M$_{\odot}$. 

\begin{figure*}
 \begin{center}
  \includegraphics[width=0.65\linewidth]{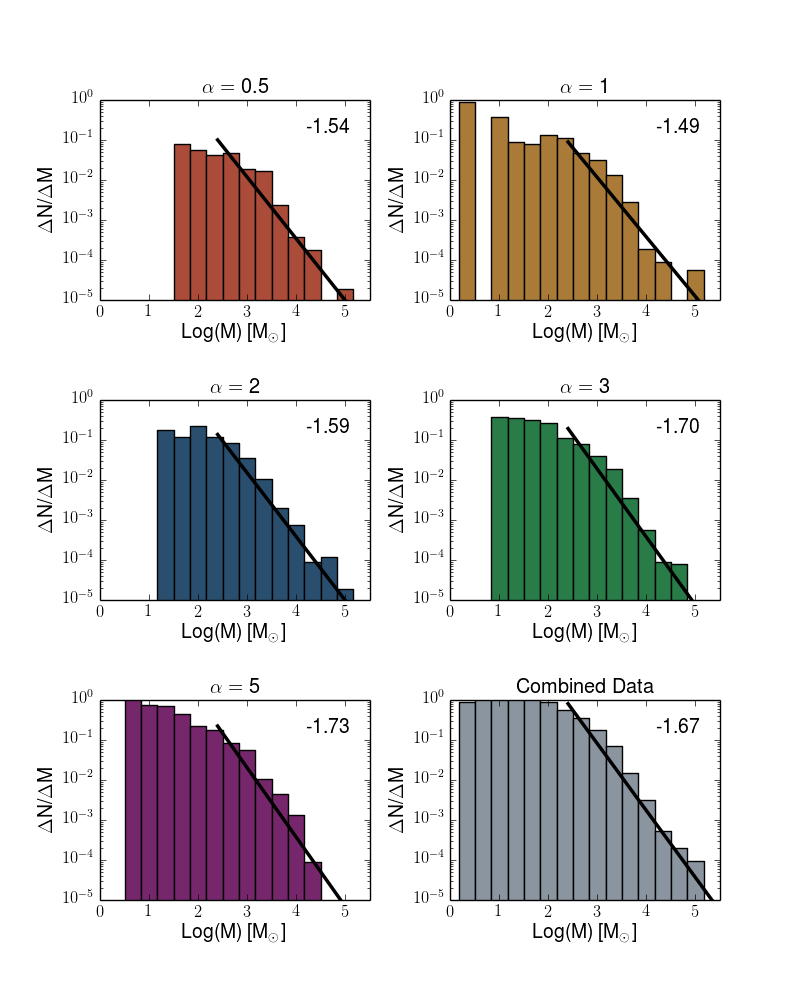}
 \end{center}
 \caption{The final cluster particle mass distributions for all simulations. The initial virial parameter is shown an the top of each panel. The final panel shows the mass function for all of the data combined. The data is fit with a straight line for masses greater than log(M) $>$ 2.4, following the work of \protect\cite{Moore}, and the resulting slope is shown in top right of each panel.}
 \label{fig:ECMF_boundedness}
\end{figure*}

\begin{figure*}
 \begin{center}
  \includegraphics[width=0.8\linewidth]{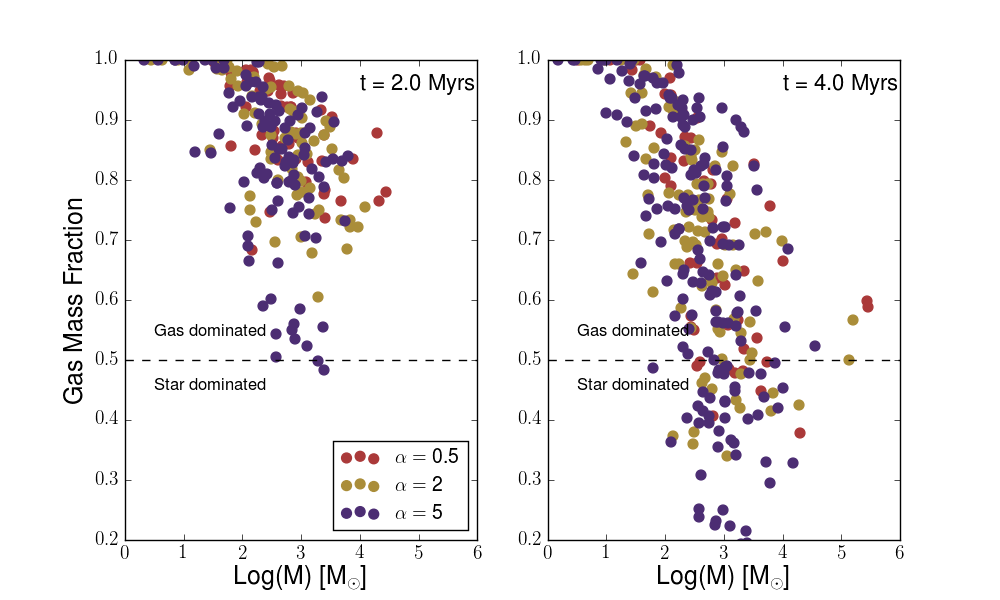}
 \end{center}
 \caption{The particle gas mass fractions, defined as the total mass in the gas reservoir divided by the total cluster particle mass, of individual particles for simulations with different initial virial parameters, shown at 2 Myr (left) and 4 Myr (right). We only show three simulations for readability. Note that the gas reservoir refers to the gas which has been incorporated into the particle, and therefore resides within the particle radius of 0.325 pc.}
 \label{fig:star_reservoir}
\end{figure*}

We compare the final total (ie. stellar content and gas reservoir) cluster particle mass distributions across simulations in Figure \ref{fig:ECMF_boundedness}. In order to compare to the results presented in \cite{Moore}, who measured the clump mass function in W43 using JCMT, we fit the distributions with a line for masses greater than log(M) $>$ 2.4 M$_{\odot}$. We discuss the observational results and the connection to our simulations in Section \ref{clumpobs}. 

The resulting slopes are shown in the upper right hand side of each panel. The fits range from -1.49 to -1.73 for simulations with an $\alpha_0$ of 1 and 5, respectively. The slope for all combined data, shown in the lower right panel, is -1.67. There is a slight trend towards steeper slopes as $\alpha_0$ increases. Since the lower $\alpha_0$ clouds have the most massive particles, more mass in cluster particles as a whole, and a smaller number of particles, this trend is easily understood. In other words, the mass distributions shift to higher masses as $\alpha_0$ decreases, which manifests as a steepening slope when fitting over a fixed range.

The mass of individual cluster particles is of two types: the cumulative mass distribution of stars as determined by generations of random sampling of the IMF from the available gas reservoirs, and the current gas reservoir. The mass in each type depends on several factors. Firstly, the age of the particle determines how long it has been forming stars suggesting that older clusters may, on average, contain a larger fraction of mass in stars relative to younger clusters. Secondly, the accretion histories of individual clusters affect the division of mass. As mentioned in Section \ref{methods}, any accreted mass is added directly to the gas reservoir. Therefore, if a cluster is undergoing strong accretion, the gas reservoir may dominate over the stellar mass.

\begin{figure*}
 \begin{center}
  \includegraphics[width=0.8\linewidth]{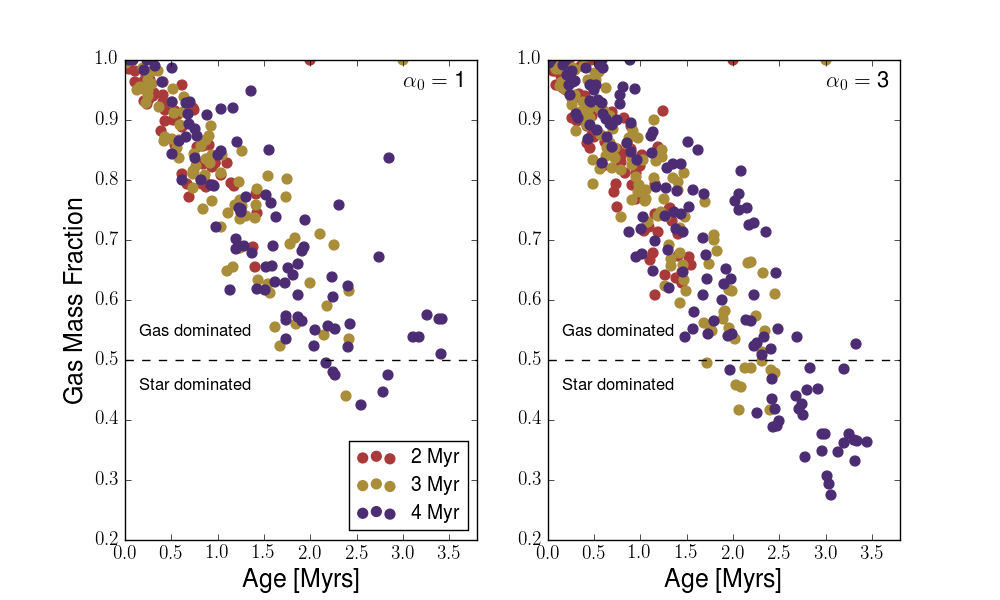}
 \end{center}
 \caption{The particle gas mass fractions as a function of cluster age for a bound ($\alpha_0$ $=$ 1) and an unbound ($\alpha_0$ $=$ 3) simulations. The colours are now used to represent different times.}
 \label{fig:reservoir_age}
\end{figure*}

We examine the division of particle mass into stars and the gas reservoir in Figure \ref{fig:star_reservoir}. We plot the gas mass fraction, defined as the ratio of the gas reservoir to the total mass, for each particle versus the total particle mass. The line at a gas mass fraction of 0.5 delineates clusters which are gas dominated from clusters which are stellar dominated. We show the gas mass fractions at 2 Myr, on the left, and 4 Myr, on the right. We only show three simulations, those with $\alpha_0$ of 0.5, 2, and 3, for readability.

Figure \ref{fig:star_reservoir} shows that the majority of cluster particles are still dominated by the gas reservoir at 4 Myr. These clusters are either undergoing strong accretion, or have not had enough time to convert their mass into stars. High gas mass fractions within clusters at late times appears to contradict observations which suggest that massive clusters are completely devoid of gas by 4 Myr \citep{Hollyhead}. Our subgrid model does not allow for gas expulsion from cluster sink particles which may explain the high gas mass fractions still present at late times. Even without gas expulsion from cluster particles, however, radiative feedback could conceivably halt further gas accretion which would naturally produce lower gas mass fractions at late times. Since this is not observed, it suggests that radiative feedback is not the only process responsible for producing gas free clusters by $\sim$4 Myr. Stellar winds \citep{Dale2008} and protostellar jets \citep{Federrath2014} may play a role in clearing young, star-forming regions of gas.

\begin{figure*}
 \begin{center}
  \includegraphics[width=0.65\linewidth]{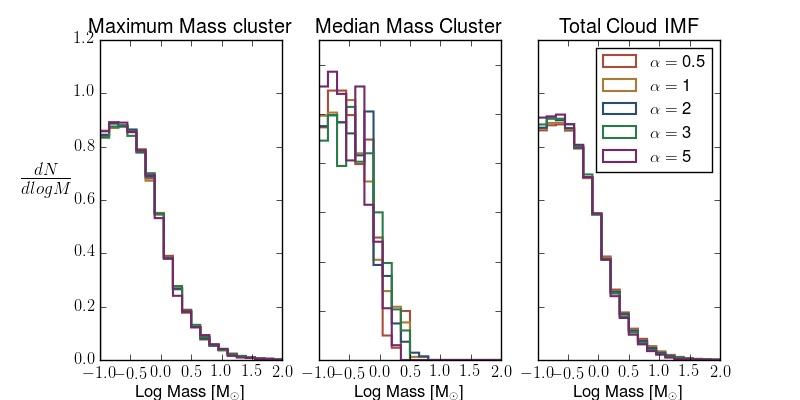}
 \end{center}
 \caption{The stellar mass distributions within the maximum mass cluster particle (left), the median mass cluster particle (middle), and the total cloud (right). See Table \ref{table:mass} for the masses of the cluster particles being shown.}
 \label{fig:IMF}
\end{figure*}

\begin{table*}
\begin{center}
\begin{tabular}{c|c|c|c|c|c|}
\cline{2-6}
 & \multicolumn{5}{ c| }{\textbf{Initial Virial Parameter ($\alpha_0$)}} \\ \cline{2-6}
 & \textbf{0.5} & \textbf{1} & \textbf{2} & \textbf{3} & \textbf{5} \\ \hline
\multicolumn{1}{ |c| }{\textbf{Maximum Total Mass (M$_{\odot}$)}} & 3.27$\times$10$^5$ & 2.22$\times$10$^5$ & 1.78$\times$10$^5$ & 1.57$\times$10$^5$ & 3.26 $\times$10$^4$ \\ \hline
\multicolumn{1}{ |c| }{ \textbf{Median Total Mass (M$_{\odot}$)}} & 368 & 485 & 313 & 290 & 318 \\ \hline
\multicolumn{1}{ |c| }{ \textbf{Maximum Stellar Mass (M$_{\odot}$)}} & 1.50$\times$10$^5$ & 1.13$\times$10$^5$ & 8.58$\times$10$^4$ & 8.70$\times$10$^4$ & 1.26$\times$10$^4$ \\ \hline
\multicolumn{1}{ |c| }{ \textbf{Median Stellar Mass (M$_{\odot}$)}} & 144 & 107 & 103 & 91 & 97 \\ \hline
\end{tabular}
\caption{Particle masses in models with varying $\alpha_0$ values.}
\label{table:mass}
\end{center}
\label{}
\end{table*}

The overall spread in gas mass fractions shows differences among the simulations. The unbound simulation (ie. virial parameter of 5) contains the largest number of stellar dominated clusters. This is likely tied to the dynamical histories of individual particles. The higher gas velocities present in the high $\alpha_0$ simulations can result in cluster particles drifting away from their local gas sources, thereby shutting off future accretion and producing stellar dominated clusters. This is less pronounced in the lower $\alpha_0$ simulations because of the global collapse of both gas and particles which facilitates ongoing accretion.

The progression from a gas to a stellar dominated cluster particle is inherently a time driven process as cluster particles convert their mass into stars. This would suggest that the oldest clusters should have the lowest gas mass fractions. This may not be case, however, because ongoing accretion can supply the clusters with a fresh gas supply. To visualize this interplay, we plot the gas mass fraction as a function of cluster age, instead of mass, in Figure \ref{fig:reservoir_age}. We show the results from two simulations, with $\alpha_0$ values of 1 and 3, to show the difference between bound and unbound simulations. We plot the gas mass fractions of all particles present at 2, 3, and 4 Myr. 

We see from Figure \ref{fig:reservoir_age} that, generally speaking, older cluster particles have lower gas mass fractions (ie. a higher fraction of stars) than younger clusters. We do, however, see significant outliers with some old clusters having high gas mass fractions. This highlights the importance of the local environment on star formation, suggesting that, even within one GMC, there can be distinct subclustered regions which can have very different star formation histories in comparison to their neighbours.

\subsection{Stellar Content}

\begin{figure*}
 \begin{center}
  \includegraphics[width=0.65\linewidth]{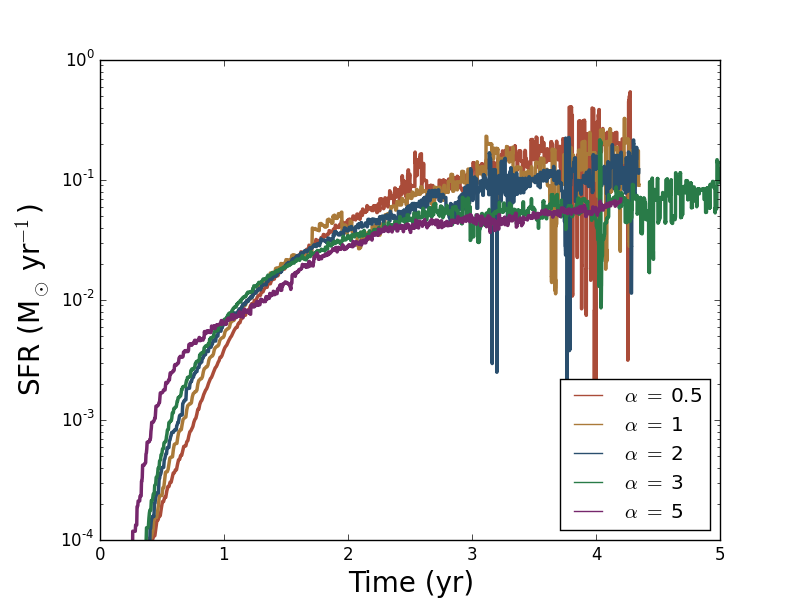}
 \end{center}
 \caption{The global star formation rates over time for the 5 simulations including radiative feedback. The curves were smoothed using a sliding average window for readability.}
 \label{fig:SFR_all}
\end{figure*}

The stellar mass distribution within individual cluster particles is partially determined by the subgrid model for star formation. A sufficiently massive particle with a large reservoir of mass to draw from will produce an IMF which is fully sampled. From \cite{Howard2014}, particles with masses of $\sim$5000 M$_{\odot}$ are sufficiently massive to form O stars. For smaller clusters, however, the amount of mass available at sampling may not be sufficient to form massive stars. 

For this reason, we examine the stellar mass distribution within individual cluster particles, as well as the total cloud, in Figure \ref{fig:IMF}. We show the stellar mass distribution in both the maximum mass cluster particle, in the left panel, and the median mass particle, in the middle panel, to illustrate the variations between different star forming regions. The right hand panel of Figure \ref{fig:IMF} shows the stellar mass distribution of the entire cloud. For reference, the mass of the the maximum and median mass cluster for all simulations is shown in Table \ref{table:mass}.

\begin{figure*}
 \begin{center}
  \includegraphics[width=0.8\linewidth]{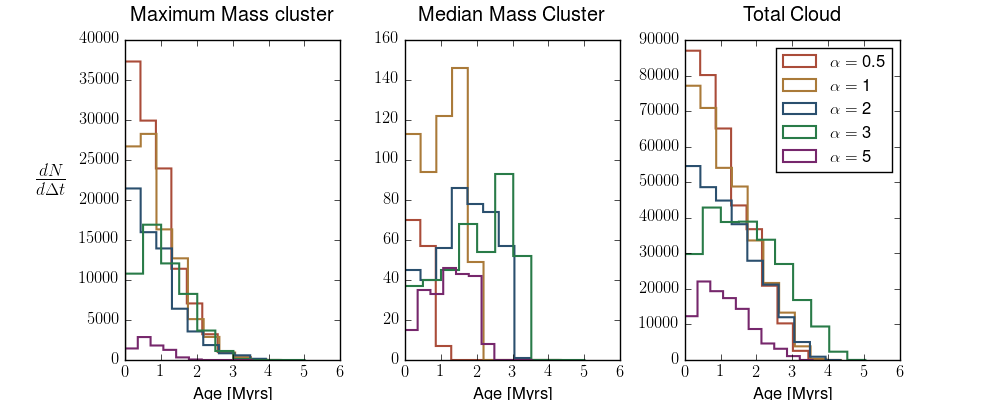}
 \end{center}
 \caption{Stellar age distributions for the maximum mass cluster particle, median mass cluster particle, and the global age distribution from left to right, respectively.}
 \label{fig:age}
\end{figure*}

As expected, the most massive particles show a fully sampled IMF with stars covering the entire mass range. There is little cluster to cluster variation among the most massive particles. We see markedly different distributions for the median mass clusters, however, which vary in mass between 290 to 485 M$_{\odot}$. Because these clusters have significantly smaller masses, they rarely form stars with masses greater than $\sim$5 M$_{\odot}$. Moreover, the distributions show stronger variations among clusters. Because these particles form a small number of stars, the effects of randomly sampling an IMF become more apparent. Overall, these distributions highlight the importance that the accretion histories play in subclustered regions and how this impacts the type of stars that are able to form.

\subsection{Star Formation Rates}

We conclude this Section with a discussion of the star formation rates within entire clouds and individual cluster particles, and relate this to the resulting stellar age spreads. 

First, in Figure \ref{fig:SFR_all}, we plot the total star formation rates from our simulations. As a consequence of our subgrid model for star formation, which only forms new stars at prescribed time intervals after a particle is formed, there are large, instantaneous jumps in the star formation rate. Since all particles do not form at the same time, the times at which particles sample the IMF are staggered. This means that, in a given timestep, there may be no particles which form new stars resulting in a SFR of zero. In other timesteps, the opposite may be true resulting in a large, instantaneous SFR. To aid in readability, the SFR curves presented in Figure \ref{fig:SFR_all} have been smoothed using a sliding average window. While this helps significantly in interpreting the Figure, there are still some spikes in the SFR that remain. 

The curves show qualitatively similar trends across simulations. There is a rapid onset of star formation, beginning around 0.5 Myr, which rises up to a roughly constant value. We do not see a turnover in the SFR. We do not expect to see a complete halting of star formation as a consequence of our subgrid model in which dense gas is assumed to remain bound within the gravitational potential well of the cluster. As shown in \cite{Howard2014}, however, the SFR will decrease for particles which are no longer accreting. This suggests that radiative feedback is not sufficient to halt the accretion onto a large number of particles and create a turn over in the SFR.

At late times, the variation in the SFR increases. This is likely due to the reason discussed above, namely that the formation of new stars in cluster particles is staggered. The variation becomes more pronounced at late times because there are more particles overall, and the particle masses increase as the simulation evolves meaning there is more mass out of which new stars can form. Additionally, the particle accretion rates become more variable at late times due to large velocities (see Figure \ref{fig:alpha}) which can either disconnect a particle from its host filament or move it to a gas rich region where strong accretion is possible. Since the SFR is intimately tied to the accretion rate of particles, this is likely contributing to the variable SFR at late times.

The final, approximately constant, SFRs cover roughly a range of 4$\times$10$^{-2}$ to 2$\times$10$^{-1}$ M$_{\odot}$ $\cdot$ yr$^{-1}$ , with variations of greater than an order of magnitude seen in some simulations. 

The evolution of the global SFR is directly tied to the resulting stellar age distributions within the simulations. We compare the age spreads in the entire cloud, as well as the maximum mass and median mass cluster particle, in Figure \ref{fig:age}. The right panel, which shows the age spread of all stars in the respective simulations, mirrors the global SFR. The majority of stars are newly formed with a rapid decline in the amount of older stars. We do not see a turn over in the distribution since star formation is still occurring at high rates.

The age distributions within the maximum mass clusters are also strongly peaked towards young ages. This is not surprising since these clusters have a significant gas reservoir to continue forming stars. The median mass clusters show more variations between simulations. Some clusters, the median cluster in the simulation $\alpha_0$ $=$ 1 for example, are newly formed and as a consequence the age distribution is narrow, $\sim$1 Myr, and peaked at young ages. The older median mass clusters, like in the simulation with $\alpha_0$ $=$ 2, show much broader distributions with age spreads up to $\sim$3 Myr which also have a turnover at intermediate ages. This shows that the formation of a stellar population in a molecular cloud is not a uniform process with distinct regions forming stars at the same time. Instead, different subcluster regions have separate star formation histories tied to their local environments.    

\begin{figure*}
 \begin{center}
  \includegraphics[width=0.8\linewidth]{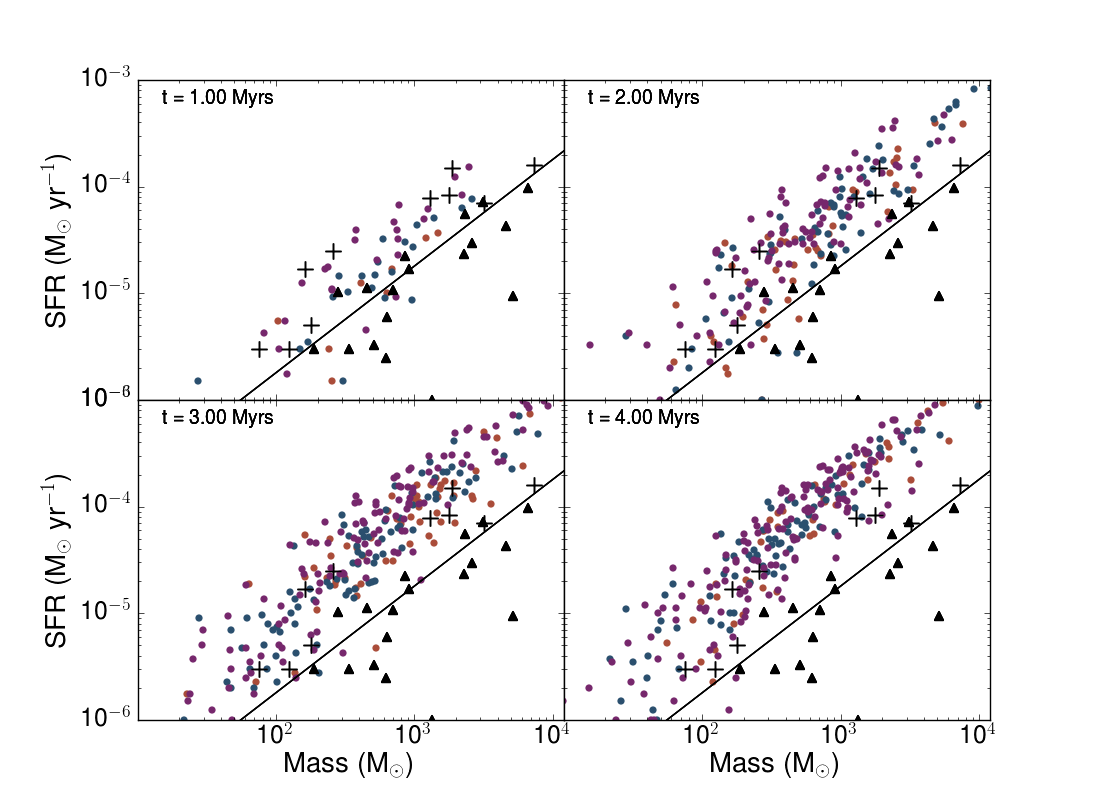}
 \end{center}
 \caption{The SFR of individual cluster particles at various times in simulations with an initial virial parameter of 0.5 (red), 2 (blue), and 5 (purple). The crosses represent the SFRs measured in local star forming regions from \protect\cite{Lada2010}. The black triangles represent another dataset presented by \protect\cite{Heidermann2010}. The solid line is an extrapolation to smaller masses from extragalactic SFR measurements performed by \protect\cite{Wu2005}.}
  \label{fig:SFRall}
 \end{figure*}

\section{Comparison to Observations} \label{clumpobs}
\subsection{Mass Distributions}

We refer back to Figure \ref{fig:ECMF_boundedness} and discuss the connection to observed clump mass function. We focus on the recent results presented in \cite{Moore} who measured masses for 1029 clumps in W43. We have chosen this dataset in particular because W43 is one of the most massive cloud complexes in the Milky Way , with a total mass of $\sim$7$\times$10$^6$ M$_{\odot}$ of which $\sim$8.4$\times$10$^6$ M$_{\odot}$ is contained in dense clumps \citep{Luong}, providing a good comparison to our simulated dataset.

The results presented in \cite{Moore} for W43 show a clump mass distribution with a high mass slope of -1.87$\pm$0.05, which is consistent with the results presented in \cite{Urquhart}. This slope is inconsistent with our results, which range from -1.49 to -1.73, and suggests that our simulations have either overproduced very massive particles, under produced low mass particles, or is a combination of both effects. There are examples of clump mass functions which have smaller slopes (eg. \cite{Gomez} and discussions therein), but these studies have different cut off masses for the fits making comparisons difficult to make. Moreover, \cite{Reid1,Reid2} measured the slope of the clump mass function in NGC 7538 and M17 and found inconsistent results between the two regions, suggesting that there may not be a universal clump mass function. As a final remark, we note that the slopes of our particle mass functions are consistent with those measured for entire GMC's by \cite{Solomon1987}, who found a slope of -1.50$\pm$0.36

The above discussion has focused on simulations which use a particle formation threshold of 10$^4$ cm$^{-3}$. As discussed in Section \ref{thresholds}, we have also completed a subset of simulations which increase and decrease the formation thresholds by an order of magnitude. In the discussion surrounding Figure \ref{fig:nstars_nthresh}, we noted that increasing the formation threshold significantly reduces the number of cluster particles while increasing the average particle mass. The will clearly impact the resulting particle mass distributions.

We find that the resulting high mass slope, analyzed in the same way as above, differs dramatically for the high formation threshold of 10$^5$ cm$^{-3}$. For an $\alpha_0$ of 0.5, the resulting slopes are -1.58, -1.54, and -1.05 for low, fiducial, and high thresholds, respectively. The corresponding slopes for the $\alpha_0$ $=$ 3 simulations are -1.81, -1.7, and -1.09. The combination of less particles and higher average masses, in the case of a high formation threshold, results in significantly shallower slopes which are less consistent with observations. This provides further justification for the choice of 10$^4$ cm$^{-3}$ as our formation threshold. Not only is this value observationally motivated, it results in particle mass distributions which are closer to measured values.

\subsection{Star Formation Rates in Molecular Clouds}

The SFRs discussed earlier were measured globally for each simulation. In order to make comparisons to observations, we plot the SFR of individual cluster particles in Figure \ref{fig:SFRall}. We again only show the simulations with an $\alpha_0$ of 0.5, in red, 2, in blue, and 5, in purple for readability. We have over plotted the observational results from \cite{Lada2010} in black crosses. These observations used 2MASS data to catalog the young stellar objects (YSOs) in 11 local star forming regions within 450 pc of the sun. The observed clouds are the Pipe Nebula, the Ophiuchus cloud, the Lupus cloud complex, Taurus, Perseus, the California clouds, RCrA, and the Orion cloud complex. How measurements of the YSO content in these clouds are used to estimate SFRs is described below.

\begin{figure*}
 \begin{center}
  \includegraphics[width=0.65\linewidth]{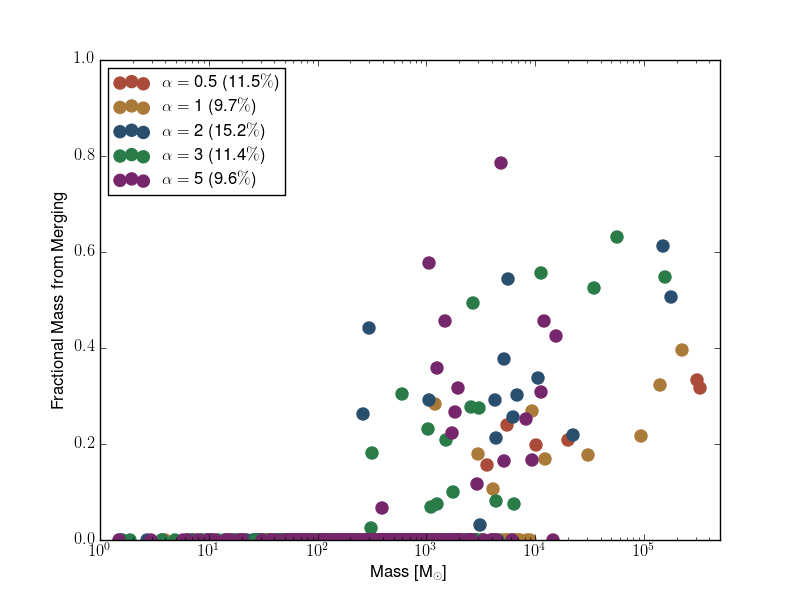}
 \end{center}
 \caption{The fraction of total cluster particle mass obtained via merging events. Clusters with fractions greater than 0.5 have grown primarily through accreting other clusters. The legend shows the percentage of surviving particles that have undergone at least one merger in its history.}
 \label{fig:merging}
\end{figure*}

The black triangles in Figure \ref{fig:SFRall} represent the dataset from \cite{Heidermann2010} who also measured SFRs in local regions, some of which are the same regions as in \cite{Lada2010}. The solid line represents the SFR-mass relation measured for extragalactic sources by \cite{Wu2005}. These observations measured the IR and HCN luminosity in the star forming cores of nearby galaxies, and used these measurements as proxies for the SFR. The resulting SFR-mass relation is given by,

\begin{equation}
SFR (M_{\odot} yr^{-1}) \sim 1.2\times 10^{-8} M_{dense}(M_{\odot})
\end{equation} 

\noindent where $M_{dense}$ is the mass of dense gas ($n$ $>$ 10$^4$ cm$^{-3}$) traced by IR emission. We have extrapolated this relation to molecular cloud masses in Figure \ref{fig:SFRall}.

In order to make a comparison to the observations, we have measured the SFR in each cluster particle in an identical manner to \cite{Lada2010}. The following formula was used to estimate the SFR in each region,

\begin{equation}
SFR = \frac{N\bar{M}}{\tau_{YSO}} = 0.25N\times10^{-6} M_{\odot} yr^{-1}
\label{SFRmass}
\end{equation}

\noindent where $N$ is the total number of young stellar objects (YSOs) in each region, $\bar{M}$ is the median mass of a YSO in M$_{\odot}$, and $\tau_{YSO}$ is the average lifetime of a YSO. The median mass, $\bar{M}$, was taken to be 0.5 M$_{\odot}$, consistent with the IMF and therefore consistent with the stellar distributions within our cluster particles. An average YSO lifetime of 2 Myr was assumed. While we do not have prestellar evolution in our subgrid model, and therefore do not know the exact number of YSOs, we do know the formation times for all stars. When solving for the SFR via equation \ref{SFRmass}, we therefore take $N$ to be the cumulative number of stars formed within the last 2 Myr to make the comparison to observations as consistent as possible. 

At early times, the SFRs agree well with the measured values, in particular the \cite{Lada2010} values. As time progresses, the SFRs in our simulations shift to higher values while following the same slope. This is consistent with the global picture presented earlier, where the global SFR rises rapidly to a roughly constant value producing age spreads which are heavily peaked at early times. While low to intermediate masses fall roughly in the range of the observations, the disparity between the observations and our results is particularly pronounced at high masses. At late times, the SFRs in the most massive cluster particles approach an order of magnitude greater than the observations. Since these massive clusters contain the most stars, this again suggests that radiative feedback is not the sole form of feedback necessary to halt the star forming process.

The high SFRs of massive clusters in our simulations are not just a product of the local environment and accretion rate, but also a product of subcluster merging which can provide these massive clusters with fresh supplies of gas to continue forming stars at high rates. 

We show this in Figure \ref{fig:merging} which plots the fraction of mass obtained via mergers. The percentage of surviving clusters which underwent at least one merging event varies from 9.6\% to 15.2\% for $\alpha_0$ values of 5 and 2, respectively, with no discernible trend as $\alpha_0$ changes. All values are shown in the legend of Figure \ref{fig:merging}. Figure \ref{fig:merging} shows that merging is not a significant form of mass accretion for the majority of cluster particles. This is particularly true for low mass clusters which grow solely via gas accretion, as shown by the large grouping of particles at zero. Above roughly 10$^3$ M$_{\odot}$, however, merging becomes a significant source of mass accretion with some particles obtaining more than half of their total mass via merging. This highlights the role that subcluster merging plays in the build up of young stellar clusters. High mass clusters in particular are not formed solely via gravitational fragmentation but rather through the merging 
of multiple, lower mass subclusters. Based on our simulations, we can conclude than subcluster merging plays a significant role in the early development of clusters with mergers producing large clusters within $\sim$ 4 Myr.

\section{Conclusions}

We have simulated a suite of turbulent, 10$^6$ M$_{\odot}$ GMCs using FLASH which have initial virial parameters in the range of 0.5 to 5. Using these, we explore the role that radiative feedback and gravitational boundedness play in the formation of star clusters and the early evolution ($<$ 5 Myr) of a GMC. To do this, we have used sink particles to represent star clusters and implemented a subgrid model to populate clusters with stars gradually over time via randomly sampling the IMF. The radiative output of these evolving clusters was then coupled to a raytracing scheme to treat the radiative transfer. The main conclusions of this work are as follows:

\begin{itemize}
\item The initial virial parameter, $\alpha_0$, strongly influences the dense clump formation efficiency, $\epsilon_{cl}$, and the star formation efficiency, $\epsilon_{sf}$, during the early evolution of the GMCs. Models that are more bound (ie. lower $\alpha_0$) have higher efficiencies in comparison to unbound models despite a delayed onset of cluster formation. This is tied to the global gas evolution of the cloud which becomes strongly centrally condensed in bound clouds, which can prolong gas accretion onto newly formed clusters, and the strong mass loss from unbound clouds. The final $\epsilon_{cl}$ values range from 37\% to 71\% and the final star formation efficiencies range from 19\% to 33\% for virial parameters of 5 and 0.5, respectively.
\linebreak
\item Radiative feedback does not strongly affect formation efficiencies over the range of time simulated. The inclusion of radiative feedback reduced $\epsilon_{cl}$ by only $\sim$1\%, except in the case of an $\alpha_0$ $=$ 3, in which $\epsilon_{cl}$ was reduced from 52\% to 41\%. This reduction was not due to radiation unbinding gas and driving outflows, but instead by suppressing cluster formation and accretion locally via heating and ionization. 
\linebreak
\item The strength of radiative feedback is enhanced if a higher formation threshold for cluster particle formation is used. We have chosen a fiducial value of 10$^4$ cm$^{-3}$ and explored using thresholds that are an order of magnitude smaller and larger than this value for a bound ($\alpha_0$ $=$ 0.5) and an unbound ($\alpha_0$ $=$ 5) cloud. In both cases, the higher threshold simulations showed larger differences in $\epsilon_{cl}$ between runs with and without feedback. A higher threshold of 10$^5$ cm$^{-3}$ for the $\alpha_0$ $=$ 0.5 model produced a difference in $\epsilon_{cl}$ between the HD and RHD simulations of 21\% compared to 2\% in the fiducial threshold simulations. The difference between the HD and RHD simulations for the $\alpha_0$ $=$ 3 model was 10\% compared to 8\% with the fiducial threshold. Since the average density of the cloud is naturally higher when clusters form in the higher threshold case, this shows that the strength of radiative feedback can be enhanced in clouds with higher density. This may 
have important implications for Globular Cluster formation which is thought to occur in high density, high pressure environments \citep{Kruijssen}.
\linebreak
\item Cluster properties are sensitive to the initial virial parameter of the cloud out of which they form. The total number of clusters formed increases as $\alpha_0$ increases, with the final number of clusters being 52 for the most bound simulation and 189 for the most unbound simulation. This, in combination with the result that bound clouds produce more massive clusters than unbound clouds, produces increasingly steeper cluster mass distributions as the initial virial parameter increases. This is a product of the mass distributions shifting to higher masses as $\alpha_0$ decreases, resulting in a steeper slope when fitting over a fixed range. The high mass slopes of our cluster mass distributions range from -1.54 to -1.73 which is only slightly shallower than the mass function presented in \cite{Moore} who measured a value of -1.87 for clumps in W43. We also find that there are more gas poor, star rich cluster particles formed in unbound simulations compared to bound simulations. While we do not have a prescription for mass loss from cluster particles, this highlights the role that dynamics plays in the early evolution of clusters. The gas poor clusters in 
bound simulations have been ejected from their host filaments effectively halting further gas accretion. In contrast, clusters formed in bound clusters are centrally condensed and continue to accrete, resulting in gas rich clusters.
\linebreak
\item We have compared the SFRs of our cluster particles to observations of local star forming regions. The SFRs in our simulations agree with observations at early times but are higher at late times, particularly for the most massive cluster particles. This is related to the global SFRs in our models which show a sharp rise in star formation which levels out to a constant value. We see no evidence for a turnover in the global SFRs which suggests that, at least in the early evolution of a GMC, radiative feedback alone is not responsible for halting star formation and dispersing the cloud. Other forms of feedback, such as protostellar jets, stellar winds, and supernovae, should be included to fully understand the cluster formation process.

\end{itemize}

We plan to extend our work by exploring different cloud properties such as magnetic field strengths and metallicity. We will also be exploring radiative feedback in cloud with masses 10$^{4-7}$ M$_{\odot}$. The highest mass clouds will probe the Supergiant Molecular Cloud regime in which Globular Cluster mass objects may form \citep{HP1994}. We will also model clouds with a range of initial densities because, as our results show, the density structure of a cloud can alter the strength of radiative feedback.  

\section*{Acknowledgments}

We thank an anonymous referee for a useful report which improved the quality of this paper. We thank Ralf Klessen and Mordecai Maclow for their invaluable input on numerics and theory, and Henrik Beuther and Jouni Kainulainen for interesting discussions of observations. C.S.H. and R.E.P gratefully acknowledge KITP, Santa Barbara, for their support and hospitality while participating as an affiliate (C.S.H) and an invited participant (R.E.P.) in the program "Gravity's Loyal Opposition: The Physics of Star Formation and Feedback". We thank Norm Murray, Andre Kravstov, and Christoph Federrath for interesting discussions during this time about simulations and theory. C.S.H. and R.E.P. also thank MPIA and the Institut f\"ur Theoretische Astrophysik (ITA) in the Zentrum f\"ur Astronomie der Universit\"at Heidelberg for their generous support during R.E.P's sabbatical leave (2015/16) and C.S.H.'s extended visit (Oct. - Nov., 2015). 

R.E.P. ~and W.E.H. ~are supported by Discovery Grants from the Natural Sciences and Engineering Research Council (NSERC) of Canada. C.S.H. acknowledges financial support provided by the Natural Sciences and Engineering Research Council (NSERC) through a Postgraduate scholarship. The FLASH code was in part developed by the DOE-supported Alliances Center for Astrophysical Thermonuclear Flashes (ASCI) at the University of Chicago. This work was made possible by the facilities of the Shared Hierarchical Academic Research Computing Network (SHARCNET: www.sharcnet.ca) and Compute/Calcul Canada.

\bibliography{howardpaper}
\bsp

\label{lastpage}

\end{document}